\documentclass{article}%
\usepackage{amsmath}%

\usepackage{authblk}

\usepackage{physics}

\usepackage{amsfonts}%
\usepackage{amssymb}%

\usepackage{todonotes}
\usepackage{graphicx}
\usepackage{caption}
\usepackage[toc,page]{appendix}
\usepackage{subcaption}
\graphicspath{{det/}}
\graphicspath{{noise/}}
\graphicspath{{transN/}}

\usepackage{dcolumn}
\usepackage{bm}
\usepackage{xcolor}

\usepackage{hyperref}

\usepackage{mathtools}
\usepackage{empheq}

\usepackage{subcaption}
\newcommand{\bx}{\mathbf{x}}
\newcommand{\by}{\mathbf{y}}
\newcommand{\bz}{\mathbf{z}}

\newcommand{\bk}{\mathbf{k}}

\newcommand{\R}{\mathbb{R}}

\providecommand{\U}[1]{\protect\rule{.1in}{.1in}}

\begin{document}

\title{Effect of Transport Noise on Kelvin–Helmholtz instability}

\author[1]{Franco Flandoli \thanks{franco.flandoli@sns.it}}
\author[1]{Silvia Morlacchi \thanks{silvia.morlacchi@sns.it}}
\author[1]{Andrea Papini \thanks{andrea.papini@sns.it}}
\affil[1]{ Scuola Normale Superiore \authorcr \normalsize\itshape P.za dei Cavalieri 7, 56126 Pisa, Italy}

\maketitle
\begin{abstract}
The effect of transport noise on a 2D fluid may depend on the space-scale of the noise. We investigate numerically the dissipation properties of very small-scale transport noise. As a test problem we consider the Kelvin-Helmholtz instability and we compare the inviscid case, the viscous one, both without noise, and the inviscid case perturbed by transport noise. We observe a partial similarity with the viscous case, namely a delay of the instability.
\end{abstract}

\section{Introduction}

Stochastic transport is a new fundamental perspective on fluid dynamics, see
e.g. \cite{Holm, Memin},~\cite{CCHMR} and \cite{FlaLuongo}. A transport type noise
in a fluid dynamic model may be seen, loosely speaking, as a simplified
description of small-medium space-scales of motion. In the numerical
simulations (see for instance~\cite{CCHMR}) we may observe the way it perturbes
large scale motion; in general, this perturbation destabilizes large scales
producing smaller eddies.

In this note we want to explore a special way transport noise may affect large
scales, somewhat opposite to the one mentioned above. The key difference is
the assumption that it is very-small-space-scale. The noise used below is made
of very small, low intensity, vortex structures. In such a case it may happen
that the transport noise acts as a dissipation, an additional viscosity. 

It corresponds to Joseph Boussinesq intuition \cite{Bouss} that ``turbulent
small scales may be be dissipative on the mean flow". The physical intuition,
beyond the specific mathematical derivation, is that fluid particles move so
erratically to produce effects similar to the molecular motion. Although a
proof is missing and the empirical validity is only moderate \cite{Schmitt},
this idea is certainly very useful for numerical simulations of fluid dynamics
model, being the basis of the LES method~\cite{Berselli}, and, in case of vorticity equations, the vortex blob method
\cite{Clements}.

Theoretically, this kind of turbulent-small-scale transport noise has been
investigated in rigorous works. This line of research was initiated
in~\cite{Galeati} and developed in several works, see e.g~\cite{FlaGalLuo,
FlaHuangPapini} and many others (see~\cite{FlaLuongo}). 

However, its stabilizing power has not been tested numerically yet. Here we
observe its action on one of the strongest and most common instabilities: the
Kelvin-Helmoltz one. We compare the inviscid case, the viscous one, both
without transport noise, and the inviscid case perturbed by transport noise.
The results are described in Section 4. Since we approximate the fluid dynamic
equations by vortex methods, which fits quite well and in a unified way with
the three cases analyzed here (inviscid, viscous and stochastic transport), we
describe some preliminaries on this topic in Sections 2 and 3. 

\section{Model formulation}

We consider the two-dimensional flow of an incompressible fluid on a 2D domain, either the full plane, $\mathbb{T}^2$ or $\mathbb{S}^2$.
As usual, the equations for the motion of the fluid are the conservation of mass and linear momentum, see e.g.~\cite{batchelor_87}, expressed through the Navier-Stokes equations in the null-divergence formulation.
The main interest for our numerical simulations is the equation for the evolution of vorticity $\omega(t,x)$, which can be derived from the Navier-Stokes equations:
\begin{equation}
\partial_t \omega + u \cdot \nabla \omega\; = \; \nu\,\Delta \omega \ .
\label{eq_nsv_eul}
\end{equation}
Here $\nu$ is the kinematic viscosity, and $u$ is the velocity of the fluid solving the NS-equation. In the 2D case, we can express $\omega := \nabla \times {u}$ as the curl of the velocity field, a vector normal to the flow plane. We denote $x=(x_1,x_2)$ to be any point in the domain.

\subsection{Point vortex method for inviscid flows}
\label{sec_vm_inviscid}

In order to state the problem we investigate numerically, we first recall several well-known facts; for a general introduction, see e.g.~\cite{MarchPulvir1, MarchPulvir2, shar}.
For the sake of simplicity, we focus on an inviscid fluid; the vorticity equation \eqref{eq_nsv_eul} reduces to,
\begin{equation}
\partial_t \omega + u \cdot \nabla \omega\; = \; 0,
\label{eq_ev_eul}
\end{equation}
From the null-divergence hypothesis, we define the stream function associated to the fluid $\psi(t,x)$, see \cite{batchelor_87}, such that the velocity $u$ is given by $u=-\nabla^\perp \psi$.
We obtain the stream function by solving the Poisson equation
\begin{equation}
\Delta\psi \; = \; - \omega.
\label{eq_lap_psi}
\end{equation}
To compute the solution, we need to express the Green's function and the convolution with the variations of constants methods, which for flows in the full plane $\mathbb{R}^2$ reads as
\begin{eqnarray}
\psi(t,x)= \int\, G(x-y)\: \omega(t,y) \; dy :=  G\ast \omega,
\label{eq_solve_psi}
\end{eqnarray}
where $G$ is the Green's function, the fundamental solution of the Laplace equation.
The explicit form of $G$ in the full plane $\mathbb{R}^2$ is
\begin{equation}
G(x,y) \; = \; \frac{1}{2 \pi}\;
{\rm \log}(|\,x-y\,|).
\end{equation}
Using \eqref{eq_solve_psi} we obtain the velocity field
\begin{eqnarray}
u(t,x)  = \int\; K(x-y)\;
\omega(t,y)\; dy :=  K\ast\omega,
\label{eq_biot_savart} 
\end{eqnarray}
where $K$ is given by
\begin{eqnarray}
K(x) & = & \nabla^\bot
G(x). \label{eq_k} 
\end{eqnarray}
The equations \eqref{eq_biot_savart} and \eqref{eq_k}
for the velocity are known as the Biot-Savart law, and $K$ is the Biot-Savart Kernel.
Note that we must correct this velocity for flows over bounded or periodic boundaries domains to satisfy the boundary condition. Therefore, we have to change the Green's function according to the Poisson equation.

Consider now a fluid particle X$_t$, moving in the velocity field; from \eqref{eq_ev_eul}, the path of the particle is
\begin{eqnarray}\label{preODE}
\frac{d}{dt}X_t & = & 
u(t,X_t)  \nonumber\\
\omega(t,X_t)&=&\omega_0(X_0).\nonumber
\end{eqnarray}
Therefore, considering the fluid as distinct ''fluid particles" of constant vorticity, these abstract objects' motion determines the scalar field's evolution. This is the premise of the point vortex method to compute \eqref{eq_ev_eul}, (see \cite{MarchPulvir1,MarchPulvir2} and \cite{shar} for details).

To this end, consider $N$ point vortices, idealizing a 2D inviscid fluid and occupying
positions $X_{t}^{1},...,X_{t}^{N}$, with intensities (circulations)
$\Gamma_{1},...,\Gamma_{N}$ respectively. They move accordingly to the following set of ordinary differential equations derived from the previous computations \eqref{preODE}:
\begin{equation}
\frac{dX_{t}^{i}}{dt}=\sum_{j\neq i}\Gamma_{j}K\left(  X_{t}^{i},X_{t}^{j}\right)
\label{ODE}
\end{equation}
where the vector-valued kernel $K\left(  x,y\right)  $ is the Biot-Savart
kernel, equal to $\frac{1}{2\pi}\frac{\left(  x-y\right)  ^{\perp}}{\left\vert
x-y\right\vert ^{2}}$ in full space, suitably modified on a torus or in a
bounded domain. One can prove that the empirical measure
\[
\omega\left(  t,\cdot\right)  :=\sum_{i=1}^{N}\Gamma_{i}\delta_{X_{t}^{i}}%
\]
is a weak solution of 2D Euler equations in vorticity form (suitably
interpreted for distributional fields as in~\cite{Choquet}):

\begin{equation}
\begin{aligned}
\partial_{t}\omega+u\cdot\nabla\omega &  =0\\
u\left(  t,x\right)   &  =\int K\left(  x,y\right)  \omega\left(  t,y\right)
dy\\
\omega|_{t=0} &  =\omega_{0}%
\end{aligned}
\label{eq:euler_w}
\end{equation}

with $\omega_{0}:=\sum_{i=1}^{N}\Gamma_{j}\delta_{X_{0}^{i}}$. Here $\omega$ is the
(scalar)\ vorticity, $u$ is the (vector) velocity.

Given a bounded probability density $\omega_{0}$, taken a sequence of i.i.d.
r.v. $X_{0}^{i}$ with density $\omega_{0}$, taken $\Gamma_{i}=\frac{1}{N}$ above,
considered now $\omega_{0}$ as a random variable with values in distributions,
the random empirical measure
\[
\omega_{N}\left(  t,\cdot\right)  :=\frac{1}{N}\sum_{i=1}^{N}\delta_{X_{t}%
^{i}}%
\]
converges weakly, in probability, to the unique solution of the above Euler
equations \eqref{eq:euler_w} (when the initial condition $\omega_{0}$ is measurable and bounded,
the Euler equations have global existence and uniqueness of bounded measurable
solutions). Similar results hold also when the approximation of the initial
condition is deterministic, and in the case when $\omega_{0}$ is more general
than a probability measure, with suitable modifications of the scheme. 

Notice that the velocity field associated to the distributional vorticity
$\omega_{N}\left(  t,\cdot\right)  $ is
\[
u_{N}\left(  t,x\right)  :=\frac{1}{N}\sum_{i=1}^{N}K\left(  x,X_{t}%
^{i}\right)  .
\]
This is a well defined vector field, of class $L_{loc}^{p}$ for every $p<2$
but not for $p=2$.

\subsection{Point vortex method for viscous flows}

To investigate viscous flows, we modifying the previous scheme by adding independent 2D Brownian motions
$W_{t}^{1},...,W_{t}^{N}$ to the equations of point vortices
\begin{equation}
dX_{t}^{i}=\sum_{j\neq i}\Gamma_{j}K\left(  X_{t}^{i},X_{t}^{j}\right)
dt+\sqrt{2\nu}dW_{t}^{i} \quad .
\label{SDE 0}
\end{equation}
Then, the empirical measure $\omega_{N}\left(  t,\cdot\right)$ converges weakly,
in probability, to the unique solution of the 2D Navier-Stokes equations in
vorticity form
\begin{equation}
\begin{aligned}
\partial_{t}\omega+u\cdot\nabla\omega &  =\nu\Delta\omega\\
u\left(  t,x\right)   &  =\int K\left(  x,y\right)  \omega\left(  t,y\right)
dy\\
\omega|_{t=0} &  =\omega_{0}%
\end{aligned}
\label{eq:euler_noise}
\end{equation}

\section{Point vortex method with environmental noise}
As mentioned above in the Introduction, several works indicated an interest in
the following stochastic modification of the Euler equations
\begin{equation}
d\omega+u\cdot\nabla\omega dt=\sum_{k\in K}\sigma_{k}\cdot\nabla\omega\circ
dB_{t}^{k}\label{SPDE}%
\end{equation}
where $\sigma_{k}=\sigma_{k}\left(  x\right)  $ are given vector fields, that
we assume divergence-free, $\left(  B_{t}^{k}\right)  _{k\in K}$ are
independent 1D Brownian motions and the stochastic operation $\circ$ stands for
the Stratonovich integral. Due to this, formally, vorticity is conserved (it is
transported randomly by the field $udt+\sum_{k\in K}\sigma_{k}dB_{t}^{k}$).\\
This model bears similarities with the viscous flows, and the objective of this paper is to show differences and similarities of the elliptic operator obtained from such a transport-advection noise.

\subsection{Transport noise and deterministic scaling limit}
The point vortex dynamics associated to the model \eqref{SPDE} is then given by the following expression:
\begin{equation}
dX_{t}^{i}=\frac{1}{N}\sum_{j\neq i}K\left(  X_{t}^{i},X_{t}^{j}\right)
dt+\sum_{k\in K}\sigma_{k}\left(  X_{t}^{i}\right)  \circ dB_{t}%
^{k}.\label{SDE}%
\end{equation}
Notice that this is a model of common noise (also called environmental noise): the BM's $B_{t}^{k}$ are the same for all particles, opposite to the model \eqref{eq:euler_noise} where each particle $X_{t}^{i}$ was affected by an independent BM $W_{t}^{i}$. See~\cite{FlaGubiPriola} for an example of theoretical results on this model. 
For models similar to this one, it has been proved (see e.g.~\cite{CoghiFla}) that the empirical measure converges to the solution of the SPDE \eqref{SPDE}.
At the same time, following~\cite{Galeati} and subsequent works, if the noise is parametrized in such a way to become more and more small scale, the SPDE \eqref{SPDE} converges to the deterministic equation with additional viscosity
\begin{equation}
\partial_{t}\omega+u\cdot\nabla\omega=\nu\Delta\omega.\label{PDE}%
\end{equation}
Inspired by~\cite{FlaLuo}, we consider a sort of mixed scaling limit: we take the point vortex dynamics with common noise \eqref{SDE}, which for given fields $\sigma_{k}$ would converge to the SPDE \eqref{SPDE}, and we choose more and more small scale coefficients $\sigma_{k}$ in order to be close to the deterministic equation \eqref{PDE}. 

The present work is numerical, but the theoretical scaling limit behind it
would be that the point vortex model \eqref{SDE} converges to the
Navier-Stokes equation \eqref{PDE}. Recalling the result mentioned in the
previous section, namely that point vortices perturbed by independent BM's \eqref{SDE 0} also converge to the Navier-Stokes equation \eqref{PDE}, we see that two different models of noise, \eqref{SDE} and \eqref{PDE}, lead to the same limit equation. 
Our aim is to explore the validity of this fact from a numerical viewpoint and in the particular case when also the coefficients $\sigma_{k}$ are point vortices.

\subsection{A digression on the theoretical selection of the noise}

In this section, in the same spirit as \cite{FlaLuo,FlaLuongo,FHP}, we explore some property of the environmental noise that we use in a simplified way  in our numerical simulations.
Following the works on modeling of passive scalars \cite{Krak}, when considering the scaling limit of [\ref{SDE}] to $\omega(\mathbf{x})$ solution of the viscous Euler equation [\ref{PDE}], we consider a model of noise in the fluid which is delta-correlated in time, namely a white noise with a precise space dependence. 
\begin{align}\label{white-noise}
\mathbf{W}\left(  t,\mathbf{x}\right)  dt=\sum_{k\in K}\mathbf{\sigma}%
_{k}\left(  \mathbf{x}\right)  dB_{t}^{k}%
\end{align}
where $(\mathbf{\sigma}_{k}\left(  \mathbf{x}\right)  )_k$ is a family of smooth divergence free vector fields on the 2D domain of the equation, and $B_{t}^{k}$ are independent one-dimensional Brownian motions; $K$ is, usually, a finite index set, but with suitable assumption we could consider also the case of countable family of smooth fields.\\
In this case, the term $\mathbf{W}\left(  t,\mathbf{x}\right)  \cdot\nabla\omega\left(\mathbf{x}\right)$ obtained in the convergence result of the point vortex empirical measure, must be interpreted as a Stratonovich integral
\[
\sum_{k\in K}\mathbf{\sigma}_{k}\left(  \mathbf{x}\right)  \cdot\nabla\omega\left(  \mathbf{x}\right)  \circ dB_{t}^{k}.
\]
This is given by an It\^{o}-Stratonovich
corrector plus an It\^{o} integral; precisely, is given by:
\[
-\frac{1}{2}\sum_{k\in K}\mathbf{\sigma}_{k}\left(  \mathbf{x}\right)
\cdot\nabla\left(  \mathbf{\sigma}_{k}\left(  \mathbf{x}\right)
\cdot\nabla\omega\left(  \mathbf{x},\mathbf{v}\right)  \right)
dt+dM\left(  t,\mathbf{x}\right)
\]
where $M\left(  t,\mathbf{x}\right)  $ is a (local) martingale. Follows that the It\^{o}-Stratonovich corrector takes the form of an elliptic operator:
\[
-\frac{1}{2}\operatorname{div}\left(  C\left(  \mathbf{x}%
,\mathbf{x}\right)  \nabla\omega\left(  \mathbf{x}\right)
\right)  dt
\]
where $C\left(  \mathbf{x},\mathbf{y}\right)  $ is  the space-covariance function of the noise
\[
C\left(  \mathbf{x},\mathbf{y}\right)  =\sum_{k\in K}\mathbf{\sigma}
_{k}\left(  \mathbf{x}\right)  \otimes\mathbf{\sigma}_{k}\left(
\mathbf{y}\right)  .
\]
As an example, we take the noise \cite{Krak}, which is relevant to our numerical investigation in the choice of the divergence-free field in the point vortex model. For simplicity, assume the domain to be $\mathbb{R}^{2}$, but modifications on $\mathbb{T}^2,\ \mathbb{S}^2$ are possible. Its covariance function is space-homogeneous, i.e. $C\left(  \bx,\by\right)  =C\left(  \bx-\by\right)  $, with the form
\[
C\left( \bz\right)  =\nu k_{0}^{\zeta}\int_{k_{0}\leq\left\vert
\bk\right\vert <k_{1}}\frac{1}{\left\vert \bk\right\vert ^{d+\zeta}}e^{i\bk\cdot
\bz}\left(  I-\frac{\bk\otimes \bk}{\left\vert \bk\right\vert ^{2}}\right)
d\bk\mathbf{.}
\]
The famous Kolmogorov 41 case follows if we take $\zeta=4/3$. Taking $k_{1}=+\infty$, then $C\left(  \mathbf 0\right)  =K\sigma^{2}$ where the constant $K$ is given by 
\[
K = \int_{1\leq\left\vert \bk\right\vert <\infty}\frac{1}{\left\vert
\bk\right\vert ^{d+\zeta}}\left(  I-\frac{\bk\otimes \bk}{\left\vert \bk\right\vert
^{2}}\right)  dk \quad .
\]
We consider small-scale turbulent velocity fields depending on a scaling parameter and taking the scaling limit in \ref{SPDE}, as in \cite{FlaLuo, Galeati}. In the case of \cite{Krak} we have
\[
k_{0}=k_{0}^{N}\rightarrow\infty
\]
The result $C\left( \mathbf 0\right)  =K\nu$ is independent of $N$, so that
the It\^{o}-Stratonovich corrector becomes equal to
\[
\nu\Delta\omega\left(  \mathbf{x}\right), 
\]
and simultaneously, we may have
that the It\^{o} term goes to zero, hence recovering \ref{PDE}.
Since this procedure can be imagined as a sequence of scaling limits with $K,N$ being large, we can't expect a precise convergence to the Laplacian operator at the level of a numerical study, but a similar effect is expected on the fluid, namely: a diffusive behaviour and a delayed formation of classical pattern and large scale structure in the fluid vorticity.

\section{Numerical results}
\subsection{Setting: Kelvin–Helmholtz instability}
In this section, we investigate classical results on the shear flow model in the setting of point vortices, analyzing the Kelvin–Helmholtz instability and the possibility of delaying the structure formation. In this way, we can both test the goodness of our point vortex models and, at the same time, set a benchmark for which we will show delayed instability.
In order to test the point vortex model in the classical cases \eqref{ODE} and \eqref{SDE 0}, we choose the particular fluid configuration of a shear flow because of its fundamental property: \textit{developing instability without viscosity and delaying it when viscosity is present}.

We work on a strip $\mathbb{S}^{2}$ equal to the set $\left[
-1,1\right]  \times\mathbb{R}$ with coordinates $x=\left(x_{1},x_{2}\right)$ and identified boundaries at $x_{1}=\pm1$; all fields are periodic in
the $x_{1}$-direction. We take an initial velocity $u_{0}$ of the form
\[
u^{0}\left(  x_{1},x_{2}\right)  =\left(  u_{1}^{0}\left(  x_{2}\right)
,0\right)
\]
and corresponding vorticity $\omega_{0}=\partial_{x_{2}}u_{1}^{0}\left(
x_{2}\right)  $. We choose, in particular, the function
\begin{equation}
u_{1}^{0}\left(  x_{2}\right)  =\left\{
\begin{array}
[c]{ccc}%
-1 & \text{if} & x_{2}\leq-\delta\\
\frac{x_{2}}{\delta} & \text{if} & -\delta\leq x_{2}\leq\delta\\
1 & \text{if} & \delta\leq x_{2}%
\end{array}
\right.
\label{eq:init_cond}\end{equation}
where we fix $\delta=0.02$ in our numerical simulations.

To compute the vorticity measure of our point vortices, we use vortex blobs, obtained by spreading the circulation of a point vortex over a chosen small area, the vortex core (see e.g.~\cite{shar}). In this formulation, the vorticity field is approximated by
\begin{equation}
\omega_\varepsilon^N(\textbf{x},t)\;=\;\sum_{i}\,\Gamma_i\,
\phi_\varepsilon(\textbf{x}-X_t^i),
\label{eq_vor_blob}
\end{equation}
where the mollifier $\phi_\varepsilon$ describes the vorticity distribution
in the vortex core, the subscript $\varepsilon$ represents the characteristic size of the vortex core.
Following standard numerical techniques (see e.g.~\cite{beale_85}), the core size $\varepsilon$ of the vortices has to be much larger than the average spacing $d$ between the vortices; the core size is usually taken to be $\varepsilon=d^q$, with $q<<1$.

In \eqref{eq_vor_blob}, the vorticity distribution at any time depends on the point vortices $X_t^i$ through the vortex blobs.
 In our numerical simulations, we take $N\sim 10^4$ point vortices; following a mean-field approach, the initial circulation for every given point vortex is derived from $u_1^0$ and is equal to $\Gamma_0^i=\frac{1}{2\delta N}$. We solved the point vortex model (\ref{ODE}) using Heun's method for a second-order time discrete approximation; the time step for our simulations was selected to be $\Delta t\sim 10^{-3}$ to ensure a trade-off between the stability of our method and the generation of vortex-like structures in the shear flow model.

In the usual way, we also recall that in our numerical framework, the kernel $K$ in \eqref{SDE} corresponds to the Biot-Savart kernel. We have that $K=\nabla^\bot G=(\partial_2 G,-\partial_1 G)$, where $G$ is the Green function on $\mathbb{S}^2$. In the whole plane we have the simple expression $G_{\mathbb{R}^2}=\frac{1}{2\pi}\log|x|$; while for our domain we know that 
\[
G(x)=\frac{1}{2\pi}\log|x|+s(x),\ \forall x\in\mathbb{S}^2\setminus\{0\},
\]
and $s(x)$ is a smooth function on $\mathbb{S}^2$. Thus, $K$ is divergence-free, smooth away from the origin, and symmetrical; moreover, it holds the following behaviour:
\[
|K(x)|\sim 1/|x|,\ \text{as}\ |x|\rightarrow 0 \quad ,
\]
which we extensively use to approximate our kernel with $K_{\R^2}$, throughout the numerical simulations.
Without ambiguity, from here on, we consider the horizontal and vertical axes as our reference frame, naming them the x-axis and y-axis, as usual.

\begin{figure}[!h]
    \centering\begin{subfigure}[t]{0.7\textwidth}
         \centering
         \includegraphics[width=\textwidth]{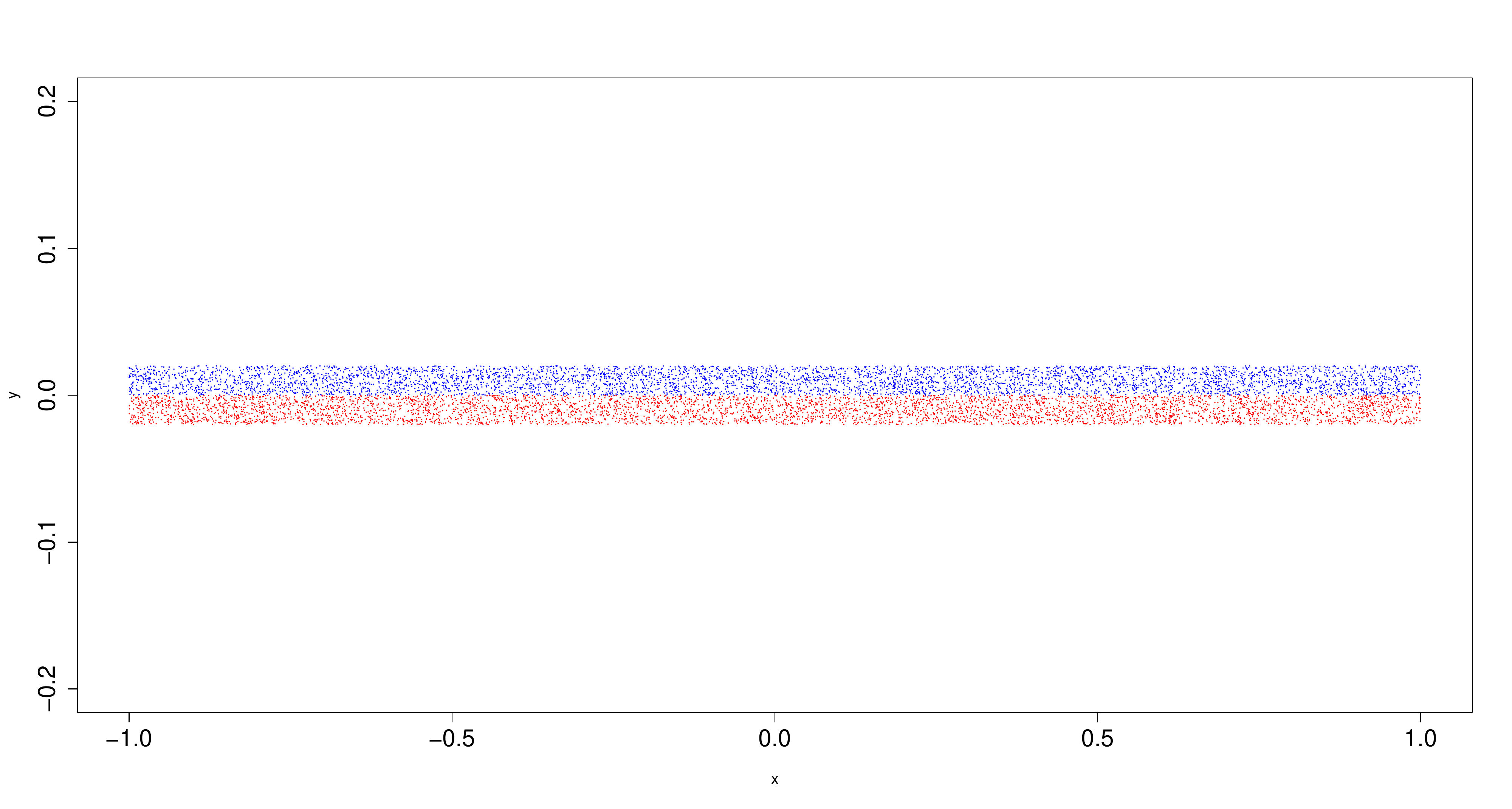}
         \caption{}
         \label{fig:y equals x 7}
     \end{subfigure}
     \begin{subfigure}[b]{0.7\textwidth}
         \centering
         \includegraphics[width=\textwidth]{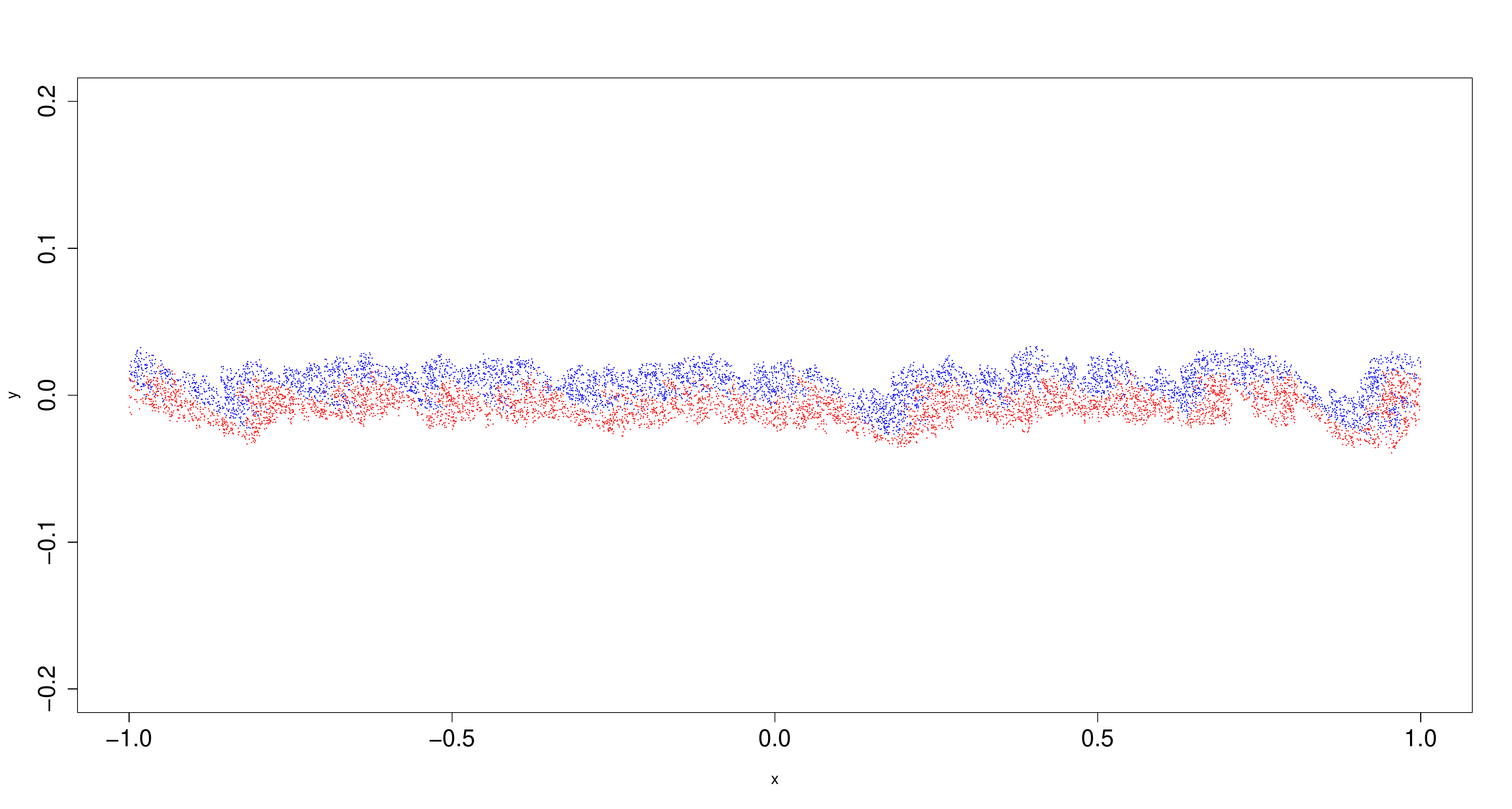}
         \caption{}
         \label{fig:three sin x 8}
     \end{subfigure}
     \begin{subfigure}[b]{0.7\textwidth}
         \centering
         \includegraphics[width=\textwidth]{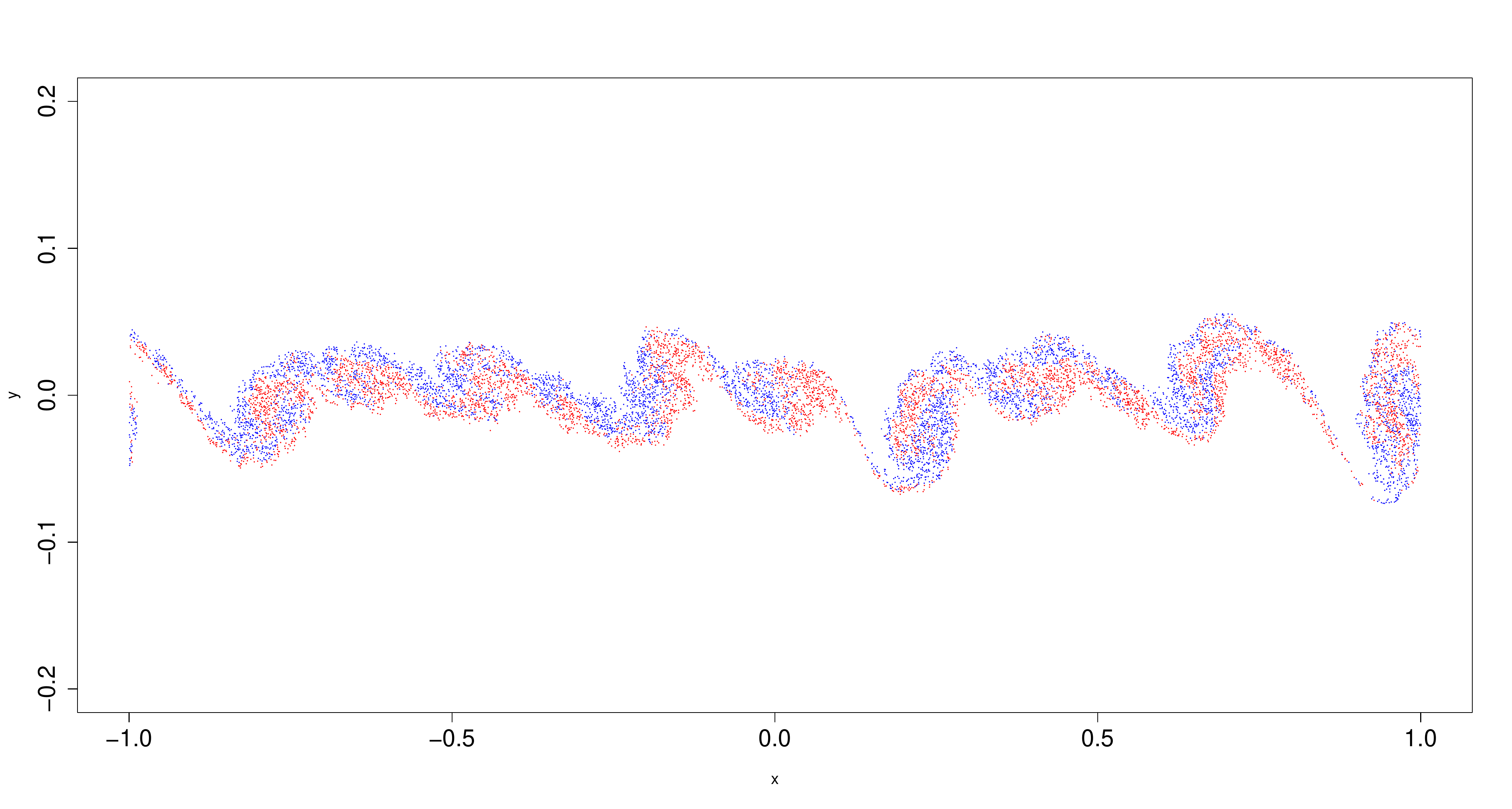}
         \caption{}
         \label{fig:three sin x 9}
     \end{subfigure}
     \caption{$\nu=0$. (a): initial configuration, approximating a shear flow fluid dynamics; (b): iteration $t=50$, formation of macroscopic vortex structures; (c): iteration $t=100$, perfectly developed macroscopic vortex structures.}
\end{figure}
\subsubsection{The role of intrinsic instability}
We know that, at least formally, the vector field $u=u^{0}$ is a solution of Euler equation \eqref{PDE} with $\nu=0$.
This system is unstable: small perturbations rapidly develop vortex blobs.
As such, we consider the system of point vortices $(X_t^i)_i$ with initial vorticity expressed as
\[
\omega_0(x_1,x_2):=\frac{1}{N}\sum_{i}\frac{1}{2\delta}\delta_{X_0^i}(x_1,x_2),
\]
where the circulation for each point is equal to $\frac{1}{2\delta N}$ and the initial positions of the vortices $X_0^i\ \forall i=1,...,N$ are uniformly distributed on the strip $[-1,1]\times[-\delta,\delta]$. The randomly generated initial condition represents small perturbations in the system and is responsible for the different pattern formation.

The measure $\omega^N_t:=\frac{1}{N}\sum_i \delta_{X_t^i}$ converges, in distribution, to the scalar vorticity field solution of the Euler equation \eqref{PDE}; analogously with the continuous case, we see in figures \ref{fig:y equals x 7} and \ref{fig:three sin x 9} the development of instability in the form of macroscopic vortex-like structure on the boundary of the two fluid layers. Note that the number of such macroscopic vortex-like structures and their position is entirely dependent on the initial condition: small perturbations on the randomly generated point vortices can produce entirely different macroscopic vortex-like structures, hence the instability of the two laminar fluids distribution.

\subsubsection{The role of viscosity and stability restoration}

\begin{figure}[!h]
    \centering
     \begin{subfigure}[b]{0.7\textwidth}
         \centering
         \includegraphics[width=\textwidth]{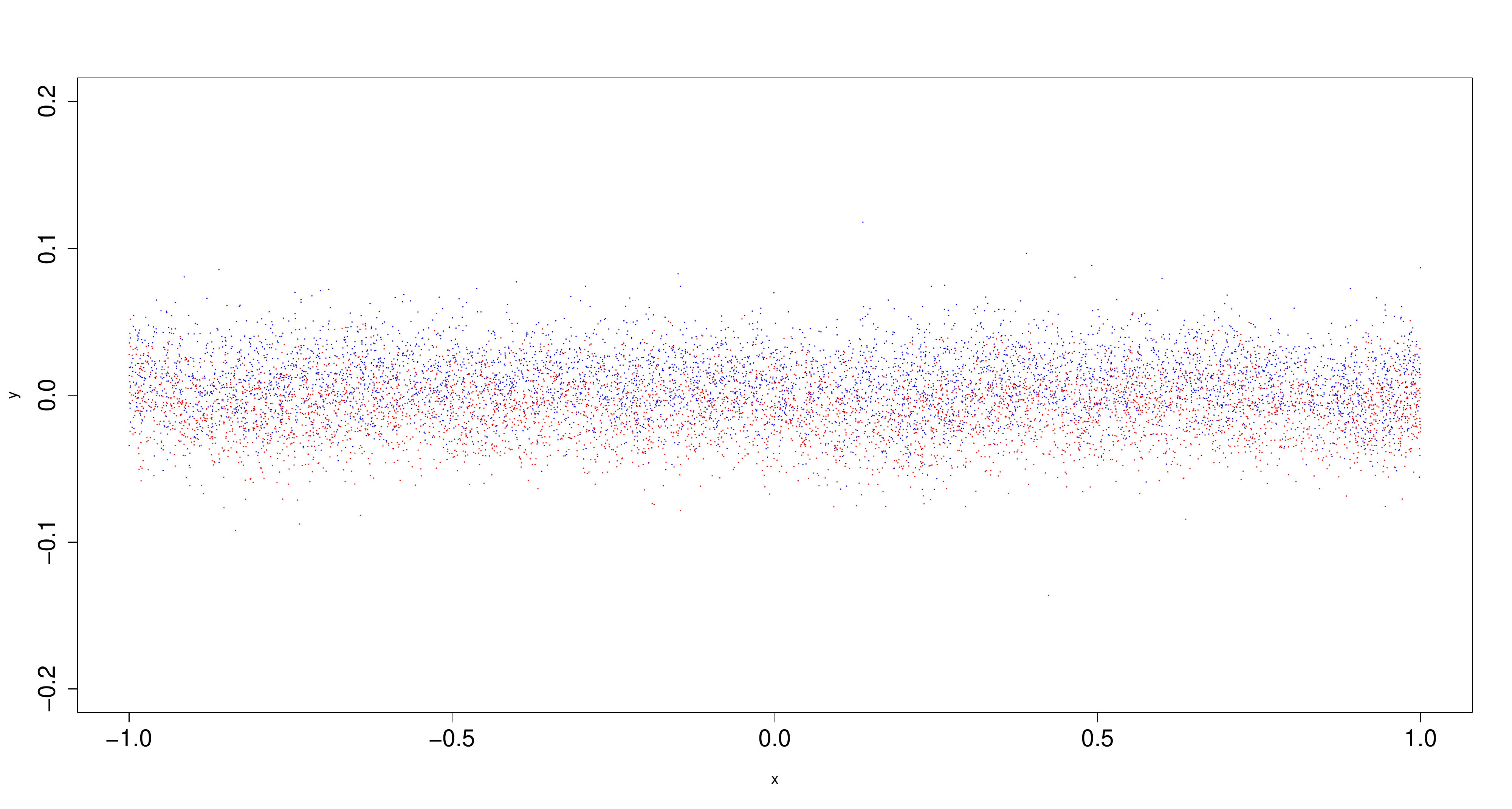}
         \caption{}
         \label{fig:three sin x 17n}
     \end{subfigure}
     \begin{subfigure}[b]{0.7\textwidth}
         \centering
         \includegraphics[width=\textwidth]{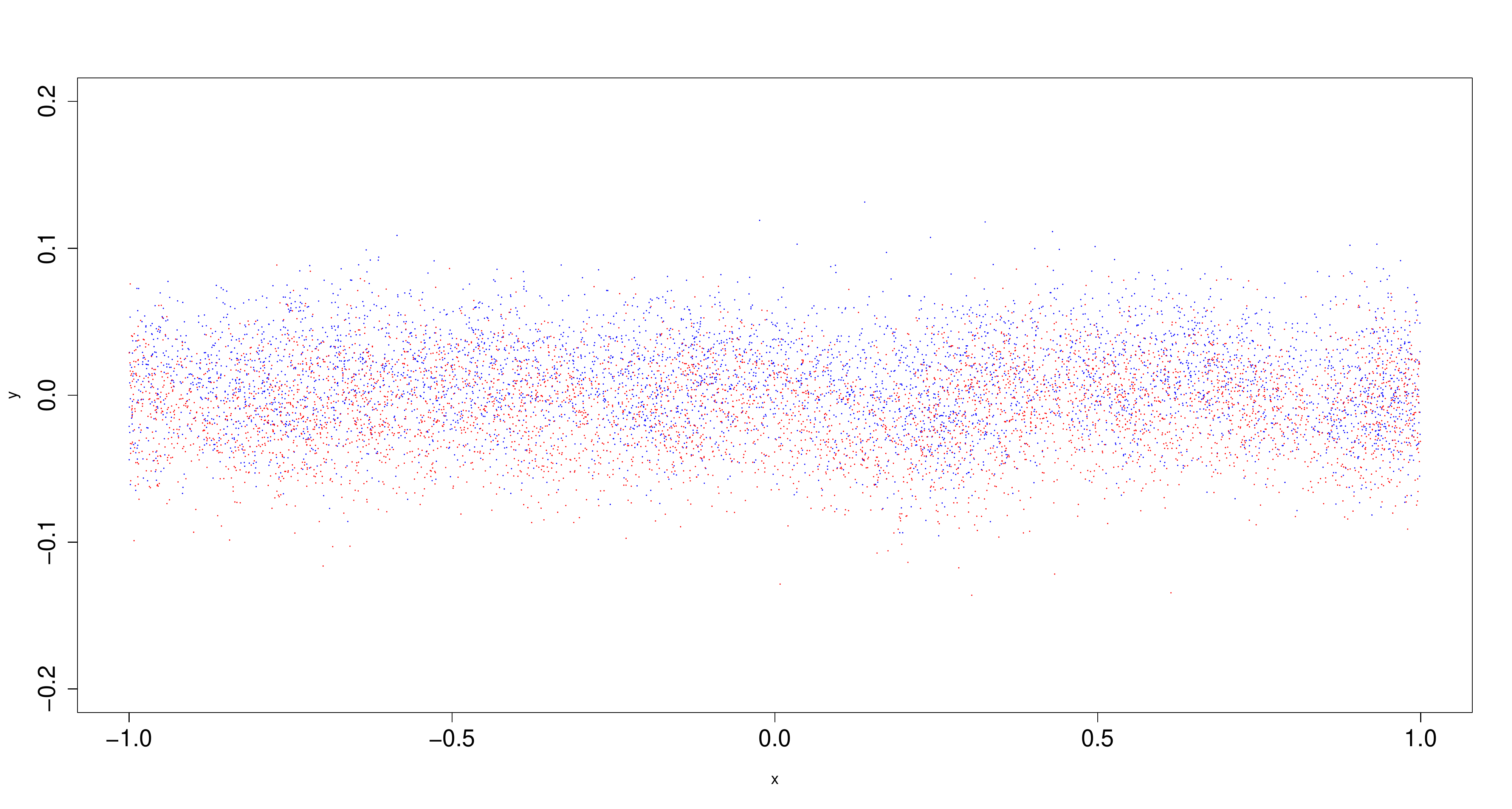}
         \caption{}
         \label{fig:three sin x 18n}
     \end{subfigure}
     \caption{$\nu>0$. (a): iteration $t=50$, preservation of strip profile; (b): iteration $t=100$, development of instability, formation of macroscopic vortex-like structures.}
\end{figure}

The exact solution of the Navier-Stokes equation \eqref{PDE}, with $\nu>0$ and
initial condition $u^{0}$, is given by
\[\label{heq}
u\left(  t,x_{1},x_{2}\right)  =\left(  u_{1}\left(  t,x_{2}\right)
,0\right) \quad, 
\]
where $u_{1}\left(  t,x_{2}\right)  $ solves the heat equation,
\begin{align*}
\partial_{t}u_{1}  & =\nu\partial_{x_{2}}^{2}u_{1}\\
u_{1}\left(  0,x_{2}\right)    & =u_{1}^{0}\left(  x_{2}\right)  .
\end{align*}
Due to the spreading of the profile $u_{1}^{0}$, the solution becomes more stable; namely, the development of vortex blobs is delayed. In our numerical simulations, we reproduce this phenomenon by perturbing the system \ref{ODE} through independent Brownian motions $B_t^i,\ i=1,..., N$, with variance linked to the viscosity parameter: $Var(B_t^i)\sim \sqrt\nu$. This system converges to the exact solution when $N\rightarrow\infty$; however, in our numerical study we are dealing with a finite system. For this reason, the profile of the strip remains quite stable for short times, with just a spread along the y-axes. We report in figures \ref{fig:three sin x 17n} and \ref{fig:three sin x 18n} a single configuration at two different timesteps, $t=50$ and $t=100$; we take $\sqrt{\nu}=0.095$ to better focus on the stability restoration.
Our results are in agreement with the theory: from a comparison with \ref{fig:three sin x 8}-\ref{fig:three sin x 9}, we see that when $\nu>0$, the profile is much more stable and diffused than in the deterministic case, and blob-like structures appear only at large times.

\subsection{Numerical results on environmental noise}

\subsubsection{Selection of divergence free field}

Starting with the same initial condition \eqref{eq:init_cond}, we consider $N+M$ point vortices, each of which we associate with a position in $\mathbb{S}^{2}$ in the following way:
\[
X_{t}^{1},...,X_{t}^{N},Y^{1},...,Y^{M}.
\]
Here, the vortices $Y^i,\ i=1,...,M$ do not move, and when activated, they represent the feedback of small-scale turbulence acting on the fluid itself on large scales.

The new simulated vortex dynamics for
$X_{t}^{i}$ as in $(\ref{SDE})$, reads %
\[\label{tnsde}
dX_{t}^{i}=\frac{1}{N}\sum_{i^{\prime}\neq i}\Gamma_{i'}K\left(  X_{t}^{i}-X_{t}%
^{i^{\prime}}\right)  dt+\sum_{j}\sigma_j(X_t^i)
\circ dW_{t}^{j}%
\]
where the Brownian motions are all independent, $B_{t}^{i}$ bi-dimensional and $W_{t}^{j}$
uni-dimensional, and they are acting simultaneously on all the particles $i=1,...,N$. The environmental noise follows the Stratonovich integral prescription, automatically implemented in Heun's method \cite{Man}.

We choose the diverge-free vector fields $\sigma_j$ as
\[
\sigma_j(X_t^i):=a_{j}^{N,M}K\left(  X_{t}^{i}-Y^{j}\right),\quad j=1,...,M
\]
following the theoretical analysis performed in \cite{FlaLuongo,FlaLuo}.  Here, the intensities $a_{j}^{N, M}$ are linked to the scaling limit process, which produces a viscosity term on the large scales, and $K$, the Biot-Savart kernel, simulates the action of such small vortices.
The idea behind such a selection is that we want to exploit the same features of the vortex model, with the formation of small-scale vortex structures generating feedback on the entire configuration.
In the limit, the dynamics of such small structures, modulated through a Brownian motion, rebound on large scales, perturbing their motion with their dissipative properties and delaying the formation of the instability.

\subsubsection{Positions and intensities of fixed vortices}

In order to make contact with previous studies~\cite{FlaLuo}, we choose the positions of the fixed vortices $Y^j$ and their intensity $a_{j}^{N,M}$ according to the convergence of the scaling limit [\ref{tnsde}].
More precisely, at each timestep, we generate $Y^j,\ j=1,...,M$, uniformly distributed point vortices; their position on the y-axis is apriori selected in the interval $[-\delta_{FX}, \delta_{FX}]$. In this setup, the vortices $Y^j$ are generated in a strip of variable height $2 \delta_{FX}$; this strip contains the moving vortices $X_t^i$, and it is taken to be of the same height of our boundary fluid layers, or one order of magnitude greater.
This choice emphasizes that our proposed ``small-scale" structures should act on all points of the fluid in all directions: the average contribution of the $Y^j$ on the $X_t^i$ along every direction should mimic a Brownian motion. We explored different setups of positions and intensity; we selected meaningful realizations, as reported in table~\ref{tab3}. 

\begin{table}[tbh]
\begin{center}
\begin{tabular}{lllll}
$M$  & $\delta_{FX}$   & $m$   & $a$      \\
200000       & 0.1 & 0.0014       &       0.0005   &           \\
132000          & 0.07 & 0.0014        & 0.0005     &          \\
1000       & 0.07 & 0.0017        &        0.005      &       
\end{tabular}
\end{center}
\caption{Parameters of the discussed realizations.}
\label{tab3}
\end{table}

We choose the intensity of the ``small-scale" perturbations following heuristic considerations. We consider the mean inter-particle distance between two fixed point vortices, $<r> =\sqrt{m}$, where $m = A/M$ is the particle density and $A$ is the total area occupied by the $M$ vortices.

Then, we focus on a single moving vortex, $X_t^i$; we compute the magnitude of its velocity when $X_t^i$ is at a distance $d =<r>/2$ from the nearest fixed vortex, i.e., its position is halfway between two fixed vortices. As a result, we obtain the following estimate for the velocity of $X_t^i$:
\[
\sum_j a_j^{N,M}\frac{1}{4\pi}\frac{|X_t^i-Y^j|^\bot}{\|X_t^i-Y^j\|^2}\sim  \frac{1}{4\pi}\sum_j \frac{a_j^{N,M}}{d}\sim \frac{a^N}{4\pi}\sum_j \frac{1}{d}.
\]
where we suppose that the coefficients depend on the configuration $(X_t^i)_i$, and are equal for each $j=1,...,M$. What is left is to estimate the number of fixed vortices such that the interaction with $X_t^i$ is not negligible: let us call this number $K$, giving us $\sim \frac{Ka^N}{4\pi d}$.

Thus, using the theory from the scaling limit of environmental transport noise (see e.g.~\cite{FlaLuongo, FlaLuo,Galeati}) and the construction of section $3.2$ for point vortices with transport noise, we assume that
\[
\nu\sim\frac{1}{2}\left(\frac{Ka^N}{4\pi d}\right)^2.
\]
This leads us to the estimate for the intensity of the fixed vortices
\[
a^N\sim 2\sqrt{2}\pi \frac{\sqrt{A}}{\sqrt{M}}\frac{\sqrt{\nu}}{K}
\]
It remains to estimate $K$, the number of the nearest fixed vortices: consider a ball centered in $X_t^i$ with radius $d$, so that the area is $A_{near}=\pi d^2$.
We recall that $m$ is the density of the fixed vortices, then the nearest vortices are: 
\[
m\times A_{near}=  \frac{M}{A}A_{near}=\frac{\pi}{4}.
\]
Taking into account only the nearest vortices, we are underestimating the actual contribution of all the vortices. In particular, we should compute such contribution by considering a radius dependent on the range of the image of the Biot-Savart kernel. For this reason, we empirically selected a wider radius $\alpha d$, with $\alpha \sim 3$. Concluding we get our estimates for the intensity
\[
a^N\sim \frac{8\sqrt{2}}{3} \sqrt{\nu}\frac{ \sqrt{A}}{\sqrt{N}}.
\]

\subsubsection{Effect of small scale common noise}

As recalled in the previous paragraph's heuristics, the procedure of the scaling limit is preserved when both $N$ and $M$ are large, and the intensity $a_j^{N, M}$ is small. For this reason, we do not search for the same exact solution of the Navier-Stokes equation \eqref{PDE}, $\nu>0$, with initial condition $u^{0}$, as per the case of the independent noise.
However, since the regime tends, in the limit, to the same solution, we expect a diffusive effect on the strip of the point vortex. More precisely, we expect to see a delay in the formation of macroscopic structures and a more dispersed displacement of such small vortex blobs that, on average, should maintain the strip configuration for a longer time.

\begin{figure}[!h]
    \centering
     \begin{subfigure}[b]{0.7\textwidth}
         \centering
         \includegraphics[width=\textwidth]{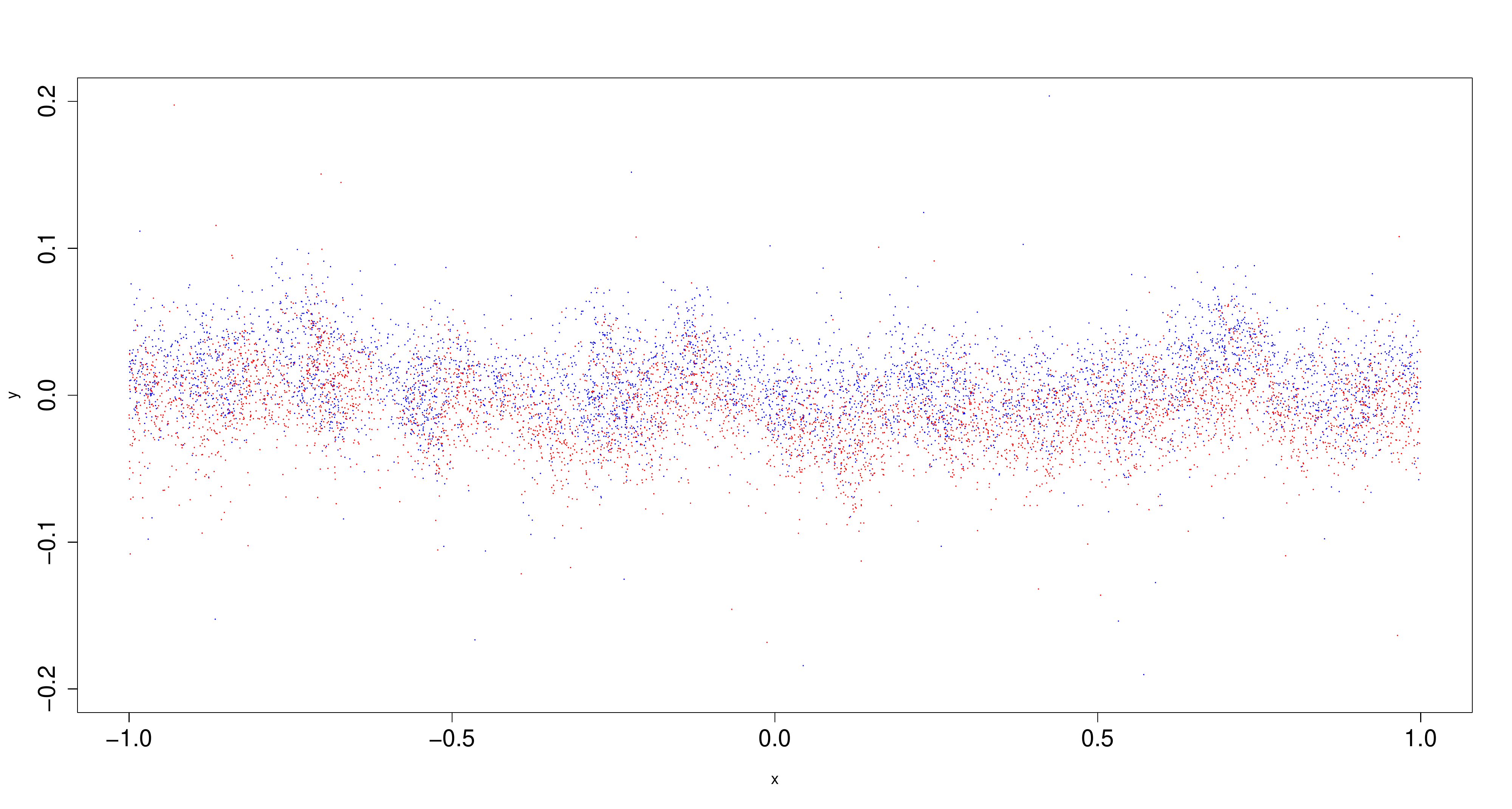}
         \caption{}
         \label{fig:y equals x17}
     \end{subfigure}
     \begin{subfigure}[b]{0.7\textwidth}
         \centering
         \includegraphics[width=\textwidth]{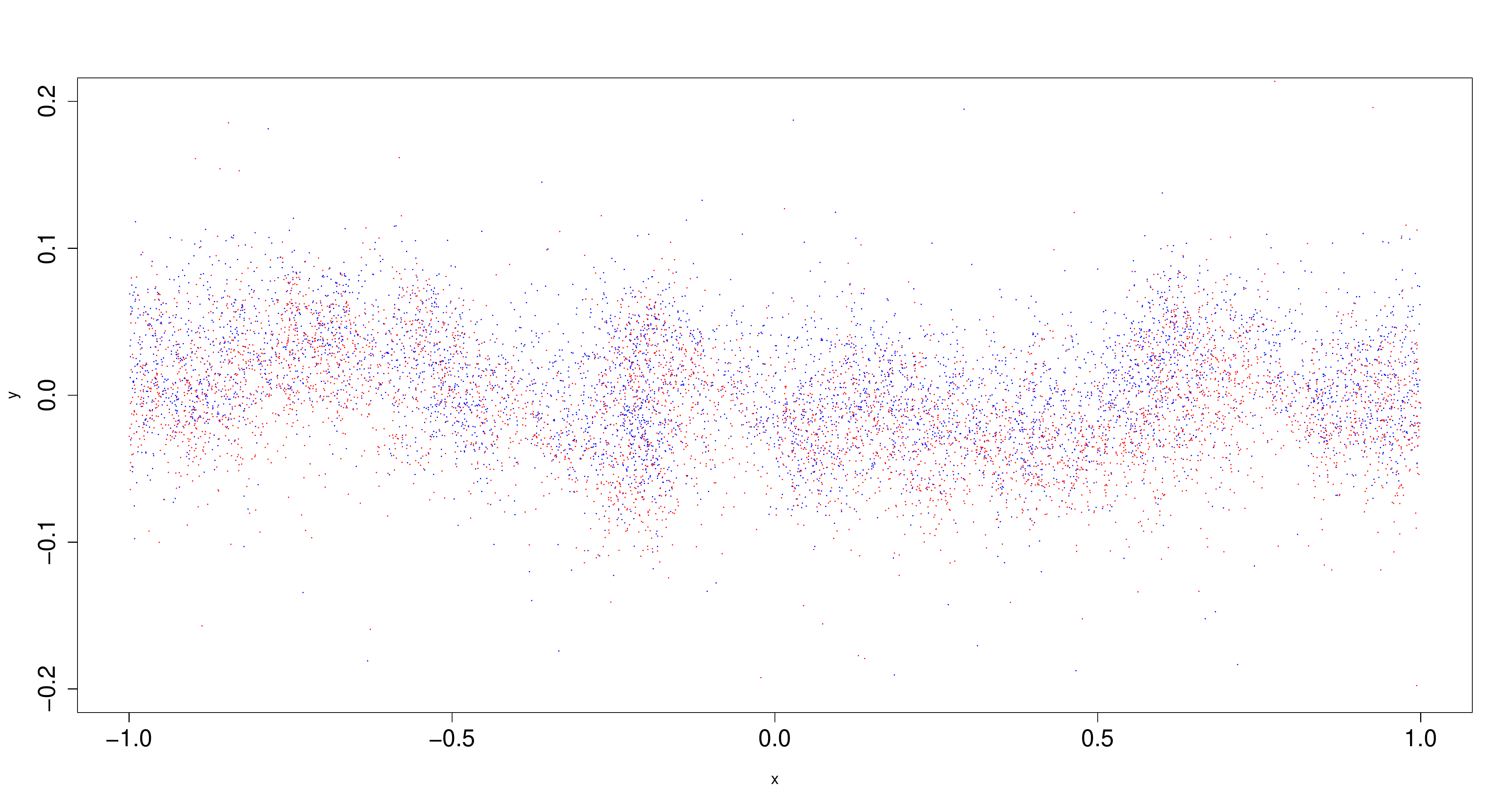}
         \caption{}
         \label{fig:three sin x18}
     \end{subfigure}
     \caption{environmental noise. (a): iteration $t=50$, diffusive behaviour of strip profile; (b): iteration $t=100$, degradation of profile, formation of macroscopic structures due to the stretch.}
\end{figure}

In the first of our simulations, we generate at each time step $M\sim 2\cdot10^5$ fixed vortices, with intensity $a_j^{N, M}\sim5\cdot10^{-4}$ which follows from the heuristics. The fixed vortices are uniformly distributed in a strip $[-1,1]\times[-0.1,0.1]$, containing the initial point vortices configuration. This particular setup captures the feedback effect of small scales on large vorticity structures, as the contribution of all the low-intensity perturbations on the dynamics of the point vortices averages in every direction.
In the proposed numerical simulation, we show that the transport noise model reproduces the desired instability delay, even if it is slightly less effective than in the independent noise case; we illustrate a snapshot of a configuration for time $t=50$ and $t=100$ in figures [\ref{fig:y equals x17},\ref{fig:three sin x18}].

From a comparison with figures [\ref{fig:three sin x 8},\ref{fig:three sin x 17n}], we see that the initial strip configuration is preserved for a longer time than in the deterministic case and rotation of the fluid is milder, but the profile is less stable than in the viscous regime. If we focus on the deterministic case, we see blob-like structures formation already at $t=50$; in contrast, in the transport noise regime, those structures are less visible and appear more prominently only at the end of our simulation ($t=100$). This delay of the instability is evident in the realizations in figures [\ref{fig:three sin x18}], compared with [\ref{fig:three sin x 9},\ref{fig:three sin x 18n}]: we notice a more diffused and homogeneous profile and a delayed formation of rotational structures due to the noise spreading the particles along the y-axis. A difference with the viscous case is that the compression in the x-axis is stronger than in the case of the independent noise, resulting in a more prominent stretch, which could resemble more the deterministic formations, placing the transport noise as a midpoint between the two regimes.

\begin{figure}[h!]
     \centering
     \begin{subfigure}[H]{0.7\textwidth}
         \centering
         \includegraphics[width=\textwidth]{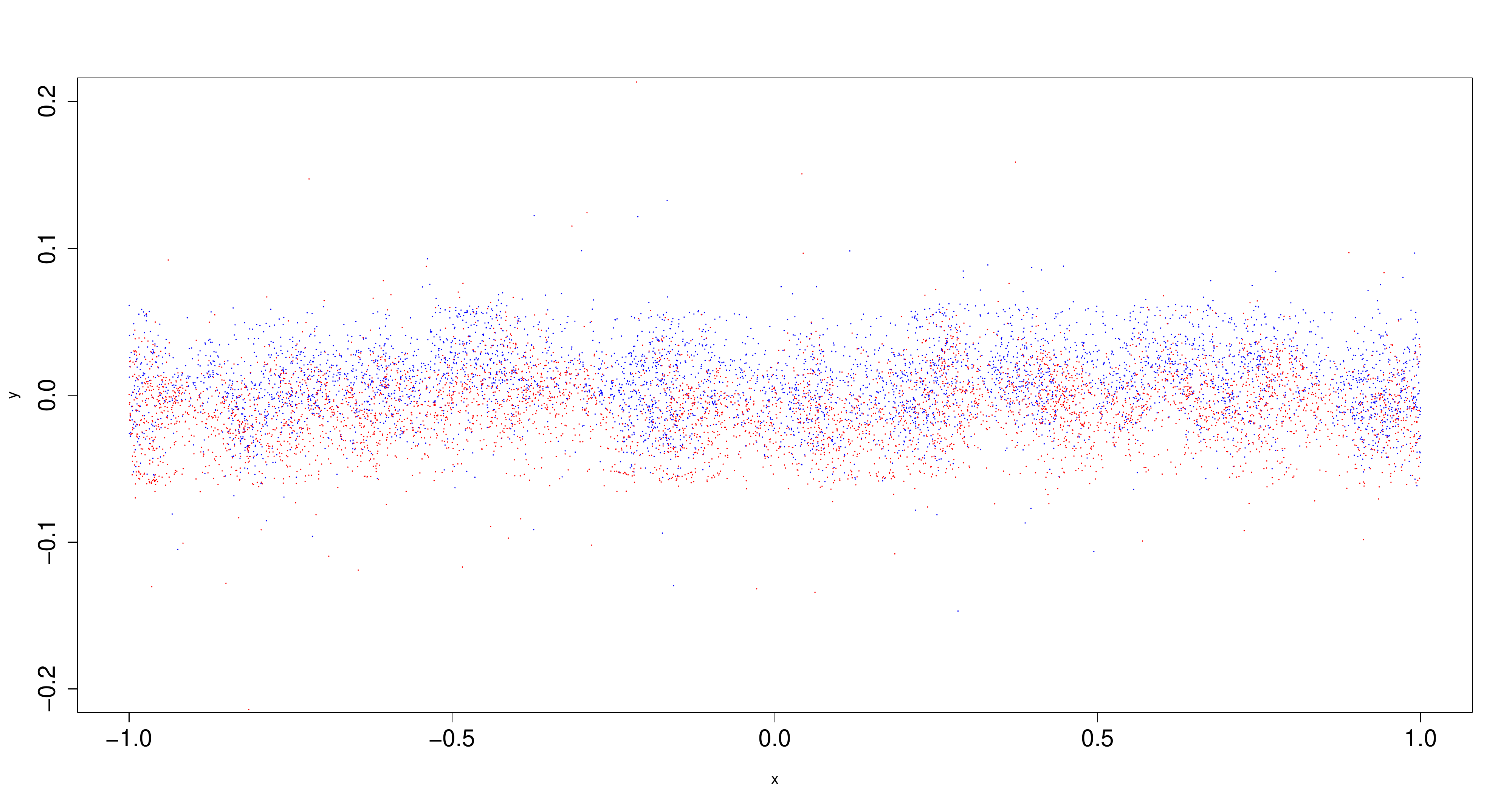}
     \end{subfigure}
     \caption{environmental noise case, iteration $t=50$, diffusion of the strip is present for a short time with preserved configuration.}
     \label{fig:y equals x131tn}
\end{figure}

In the second of our simulations, the strip of fixed points $Y^j$ is generated in the same region as the point vortices at each timestep; we selected $M\sim 1,32\cdot10^5$, the fixed vortices are uniformly distributed in $[-1,1]\times[-\delta-\epsilon,\delta+\epsilon]$, with $\epsilon=0.03$, and their intensity is derived from the heuristics $a_j^{N, M}\sim5\cdot10^{-4}$.
The results are shown in figure [\ref{fig:y equals x131tn}]: diffusion on the y-axis is present for short times, and preservation of the strip profile is guaranteed. However, the drawback of such a configuration is that analysis can be performed only for a short time: boundary effects of the fixed vortex strip can deteriorate the configuration, making the results unrealistic. In future works, we expect to overcome this obstacle by proposing a new method, now in the study, to generate small vortices only in regions activated by the shear flow's movement.

We performed a final simulation, in which we take the density of fixed vortices to be smaller than the density of the point vortices. 
\begin{figure}[h!]
     \centering
     \begin{subfigure}[H]{0.7\textwidth}
         \centering
         \includegraphics[width=\textwidth]{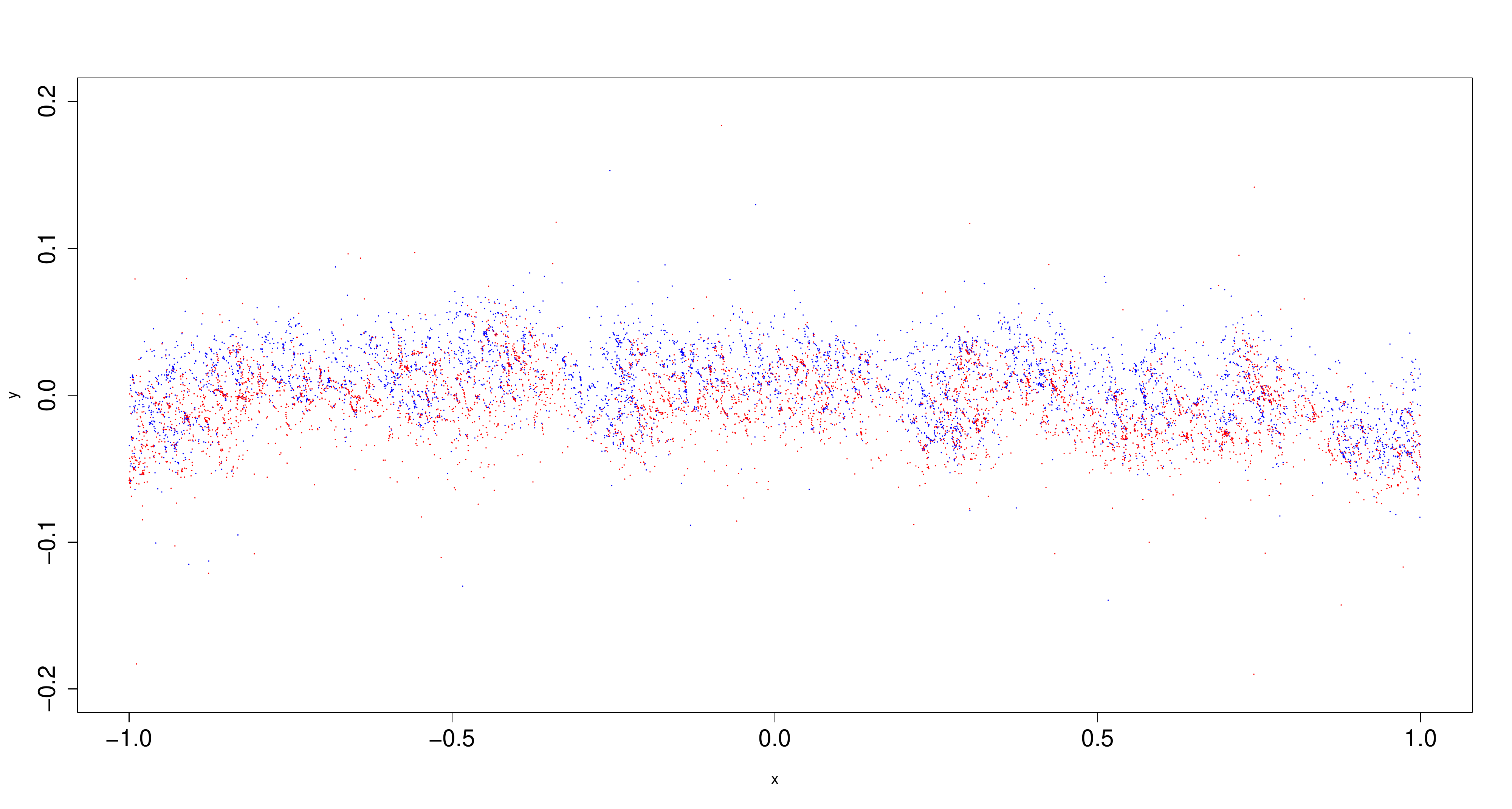}
     \end{subfigure}
     \caption{environmental noise case, iteration $t=50$, low density ratio between fixed vortices and point vortices showing emergence of medium-scale structures.}
     \label{smallx131tn}
\end{figure}
In particular, the strip of fixed points $Y^j$ is generated in the same region as the point vortices at each timestep; we selected $M\sim 10^3$, the fixed vortices are uniformly distributed in $[-1,1]\times[-\delta-\epsilon,\delta+\epsilon]$, with $\epsilon=0.05$, and their intensity derived from the heuristics $a_j^{N, M}\sim5\cdot10^{-3}$. 
While the diffusive behaviour is lost, as shown in figure [\ref{smallx131tn}], the strip is already broken at time $t=50$, showing rotating structures. The fixed vortices' low density and higher intensity seem to produce new formations and medium-scale structures, showing a completely different behaviour than the deterministic and viscose counterparts. For this reason, we need to investigate further the link between the ratio of vortices densities and the formation of new independent medium structures.

\subsection{Diagnostics}
In this section, we perform stochastic analysis on the three configurations proposed in this study to highlight the differences and the reconstructed stability, or the emergence of new structures, in the Kelvin-Helmholtz instability problem.

\begin{figure}[h!]
    \centering
     \begin{subfigure}[H]{0.49\textwidth}
         \centering
         \includegraphics[width=\textwidth]{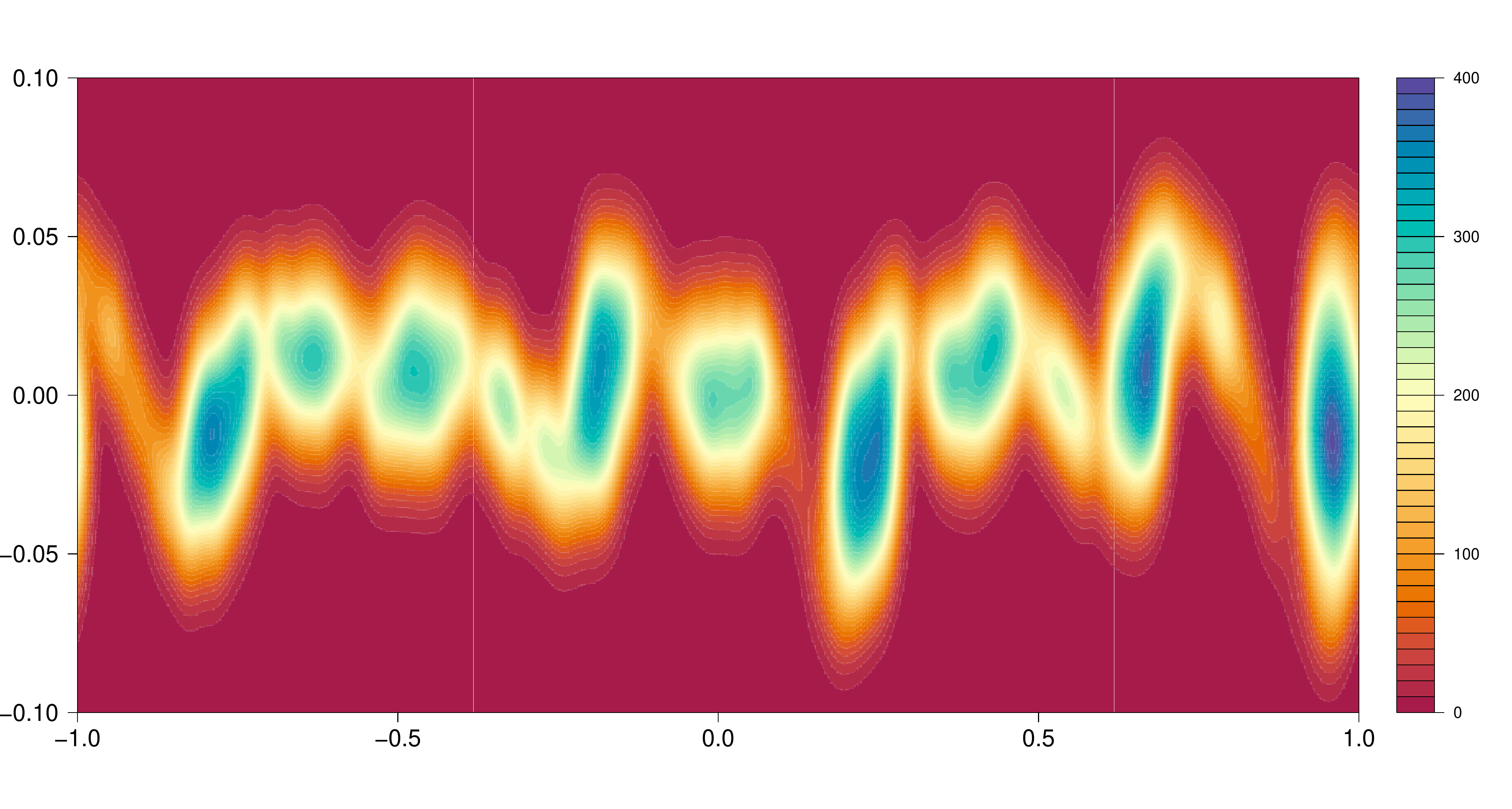}
         \caption{}
         \label{vd}
     \end{subfigure}
     \begin{subfigure}[H]{0.49\textwidth}
         \centering
         \includegraphics[width=\textwidth]{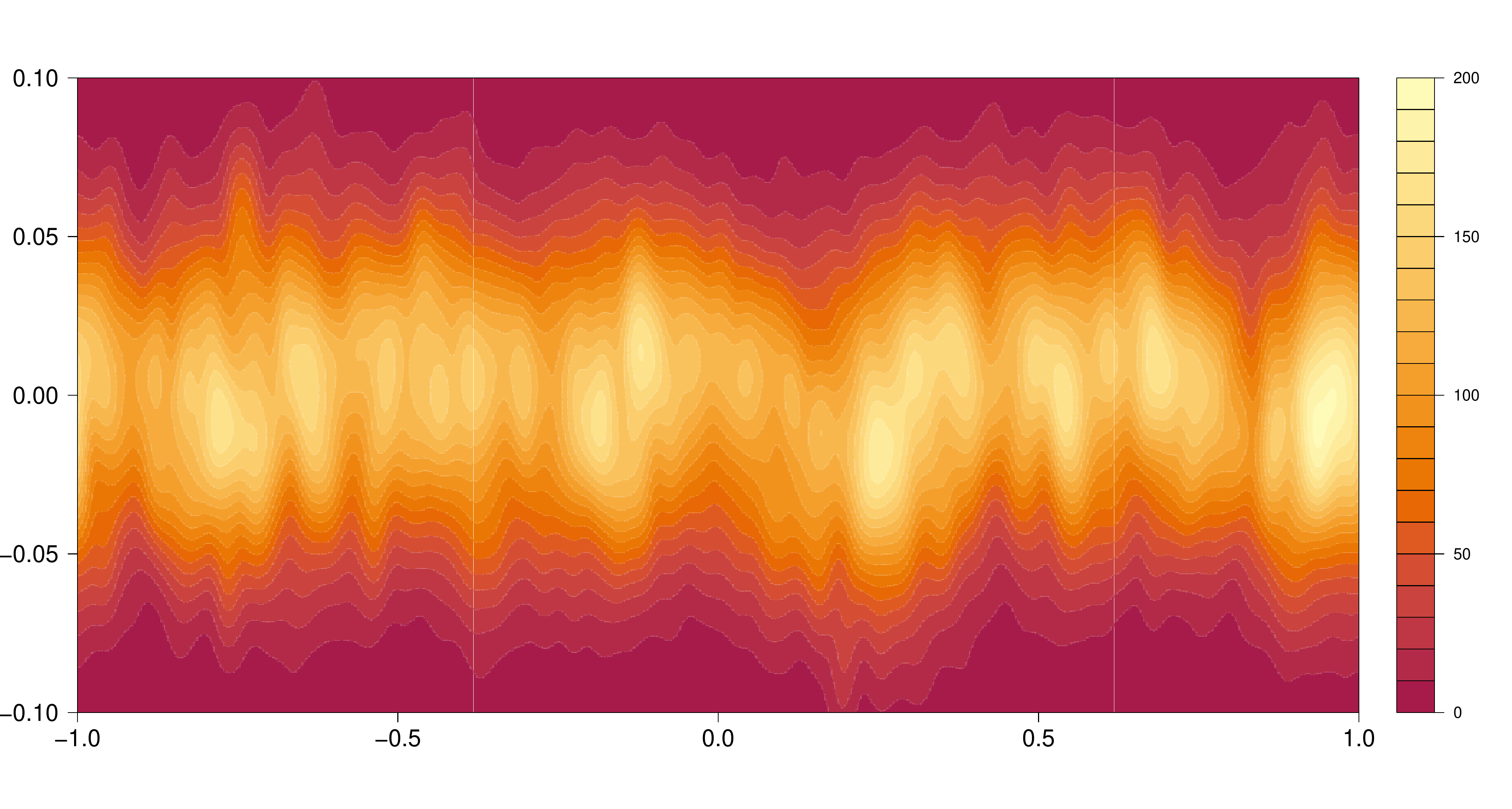}
         \caption{}
         \label{vn}
     \end{subfigure}
     \begin{subfigure}[H]{0.49\textwidth}
         \centering
         \includegraphics[width=\textwidth]{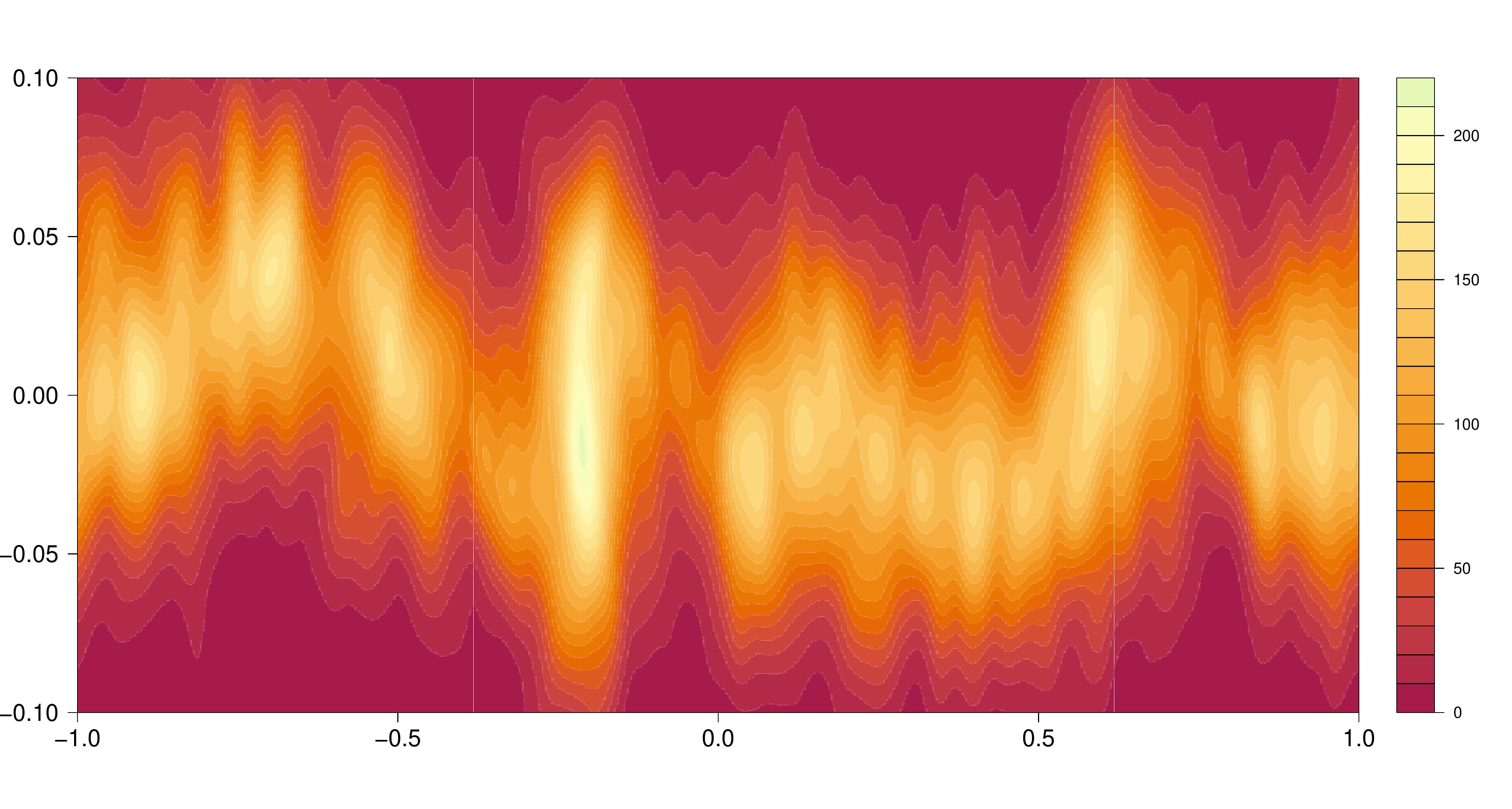}
         \caption{}
         \label{vtn}
     \end{subfigure}
     \caption{vorticity $\omega_\epsilon$ at $t=100$ in the case (a) inviscid, (b) viscous and (c) transport noise, showing macroscopic structures formation or delay of instability.}
     \label{vx}
\end{figure}

As a first step, we compute the vorticity $\omega_\epsilon$ obtained through the same mollification as in Majda \cite{beale_85}, through the vortex blob method applied to each of the point vortices $X_t^i$. 
We report our results for the vorticity computed in the deterministic case at $t=100$ in figure [\ref{vd}]. We see that the vorticity measure concentration near the fluid's boundary layer is located in the newly developed macroscopic structures. Moreover, a displacement from the initial configuration where the laminar fluid started its evolution is present.

In contrast, the vortex blob solution with viscosity $\nu>0$ retains its structure for longer times than in the inviscid case. The instability delay is graphically evident both from the configuration reported in figure [\ref{fig:three sin x 17n}] and the vorticity intensity reported in figure [\ref{vn}]: the vorticity measure concentration at time $t=100$ is similar to the one of the initial strip but with a more diffused profile on the horizontal line.

Finally, in figure [\ref{vtn}], we show the vorticity in the environmental noise regime at $t=100$. In contrast with the inviscid case, we see no development of macroscopic structures in the profile. However, even though a more diffused profile, with lower density overall, is present, the stability is lost at larger times. 
This instability at larger times suggests that different behavior, dependent on the density of fixed vortex $Y^j$ and selection of transport noise fields $\sigma_j$, could arise in applying this kind of small-scale approximation. For this reason, we focus our analysis on small-time behaviour.

\begin{figure}[!h]
    \centering
     \begin{subfigure}[H]{0.49\textwidth}
         \centering
         \includegraphics[width=\textwidth]{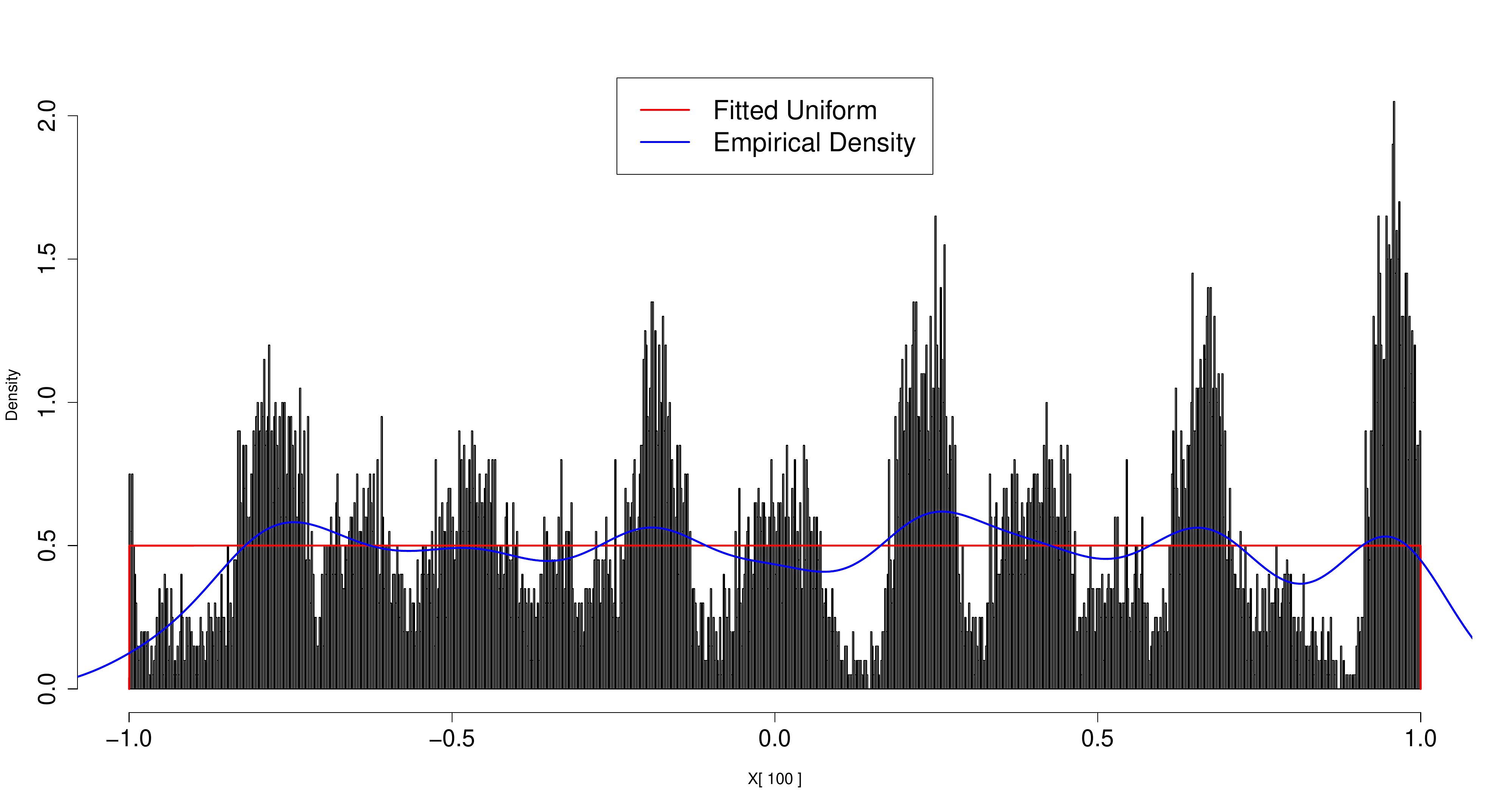}
         \caption{}
         \label{hxd}
     \end{subfigure}
     \begin{subfigure}[H]{0.49\textwidth}
         \centering
         \includegraphics[width=\textwidth]{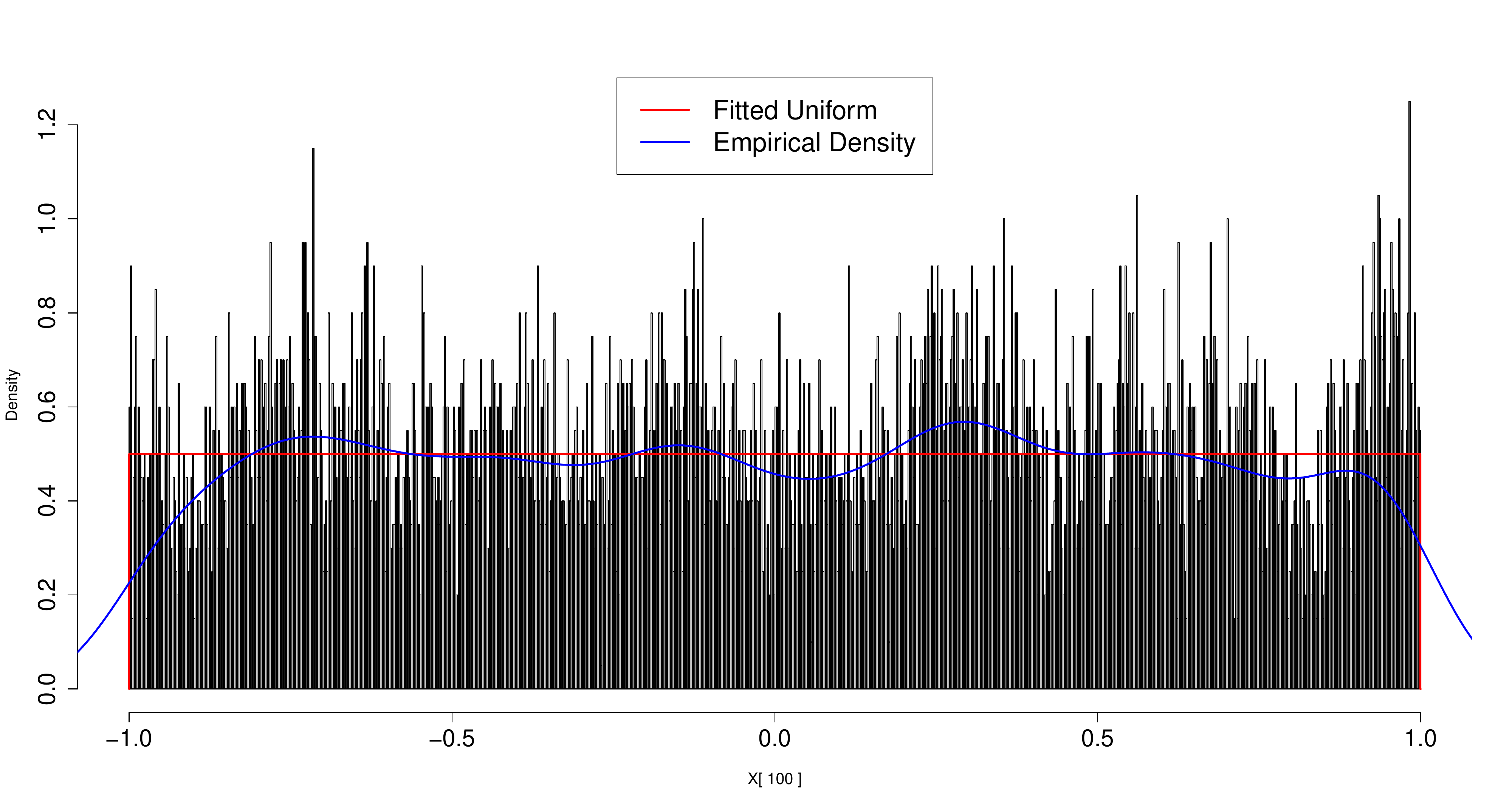}
         \caption{}
         \label{hxn}
     \end{subfigure}
     \begin{subfigure}[H]{0.49\textwidth}
         \centering
         \includegraphics[width=\textwidth]{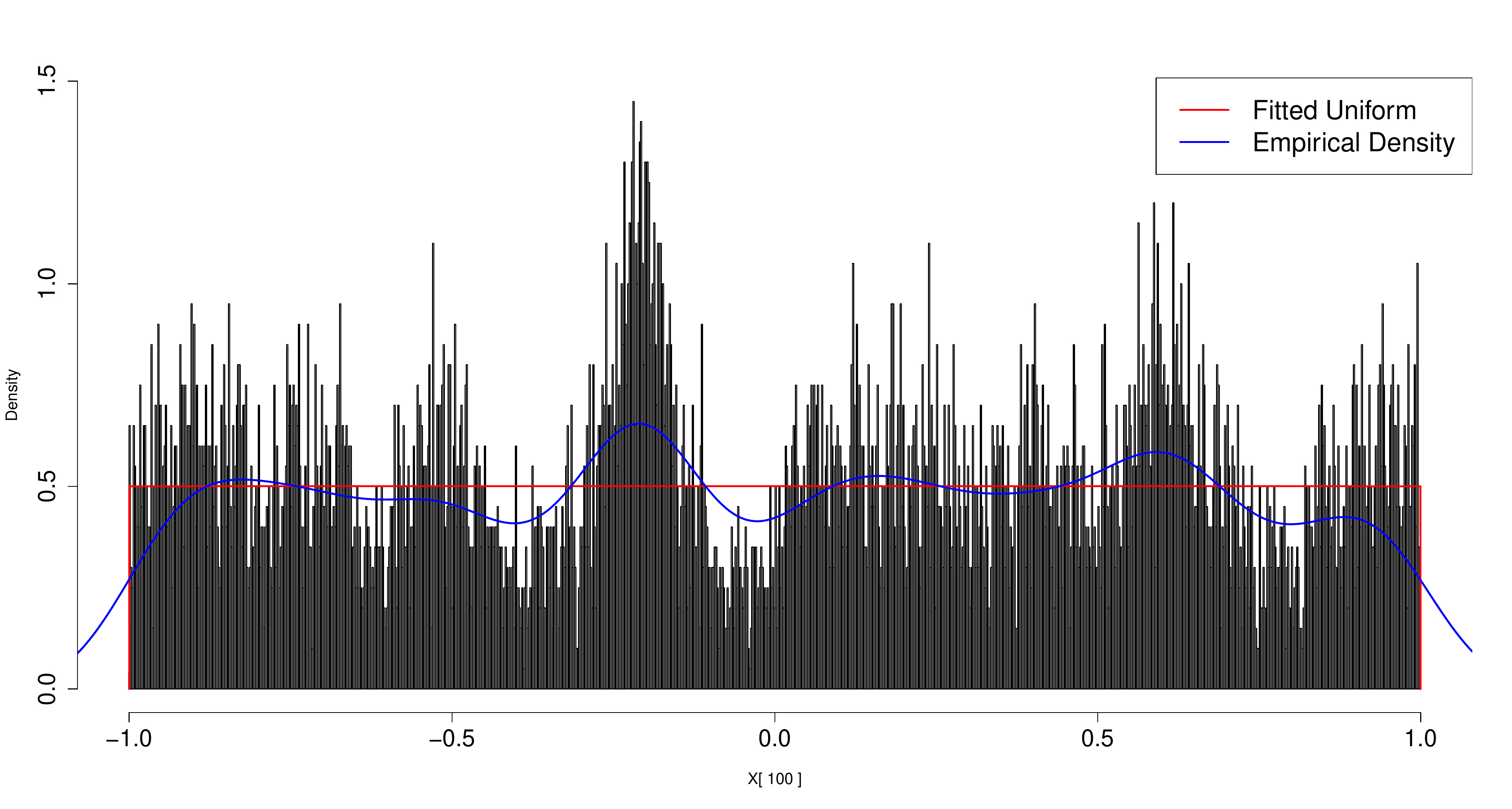}
         \caption{}
         \label{hxtn}
     \end{subfigure}
     \caption{histograms of x-positions and empirical density at $t=100$ in the case (a) inviscid, (b) viscous and (c) transport noise, showing formation of macroscopic structures.}
     \label{hx}
\end{figure}
Concerning the formation of large rotating structures, we see that the particles spread in the horizontal direction when a forcing term, either an independent or transport noise, acts on the fluid, in contrast to the solution of the Euler equation with $\nu=0$. In particular, we focus on the empirical density obtained from the x-axis in the three configurations at $t=100$, [\ref{hx}]. The deterministic case [\ref{hxd}] shows a complete formation of separate blobs with peaks in the exact locations of the macroscopic structures, as in \ref{fig:three sin x 9}. On the contrary, in the viscous \ref{hxn} and transport noise case \ref{hxtn}, the distribution of the vortices is more uniform, and it delays the instability of the fluid layers.

\begin{figure}[h!]
    \centering
     \begin{subfigure}[H]{0.49\textwidth}
         \centering
         \includegraphics[width=\textwidth]{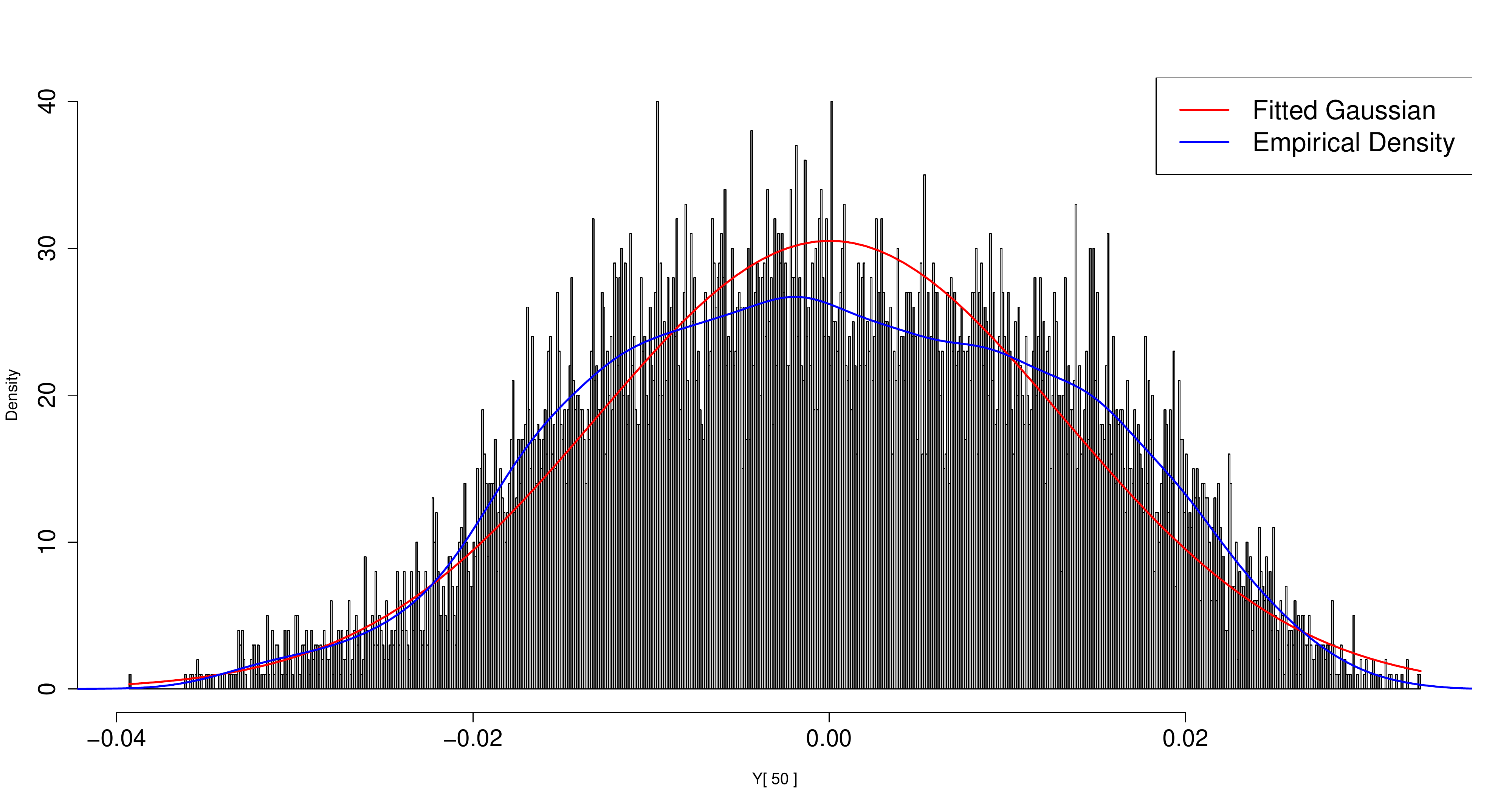}
         \caption{}
         \label{hyd}
     \end{subfigure}
     \begin{subfigure}[H]{0.49\textwidth}
         \centering
         \includegraphics[width=\textwidth]{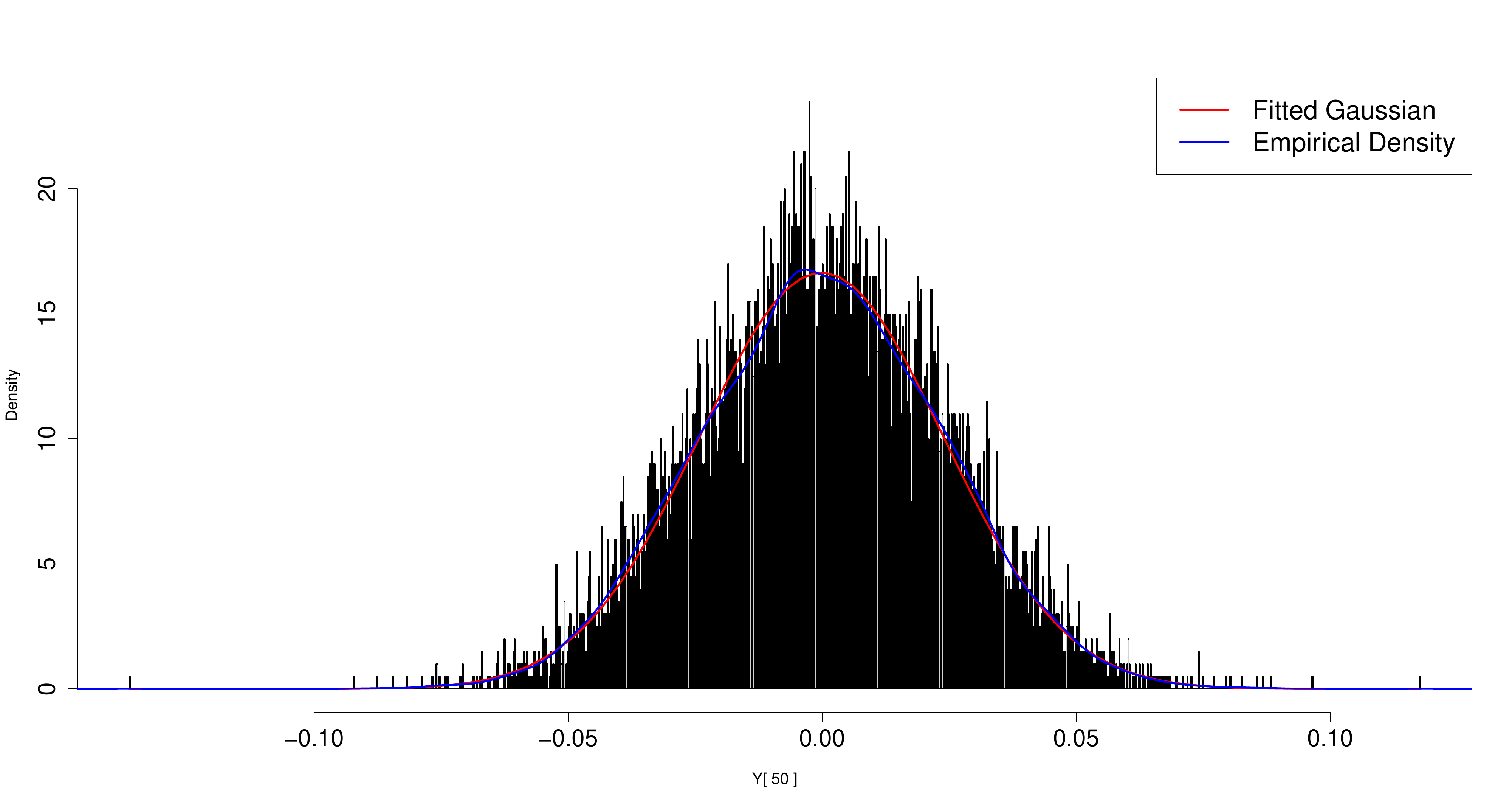}
         \caption{}
         \label{hyn}
     \end{subfigure}
     \begin{subfigure}[H]{0.49\textwidth}
         \centering
         \includegraphics[width=\textwidth]{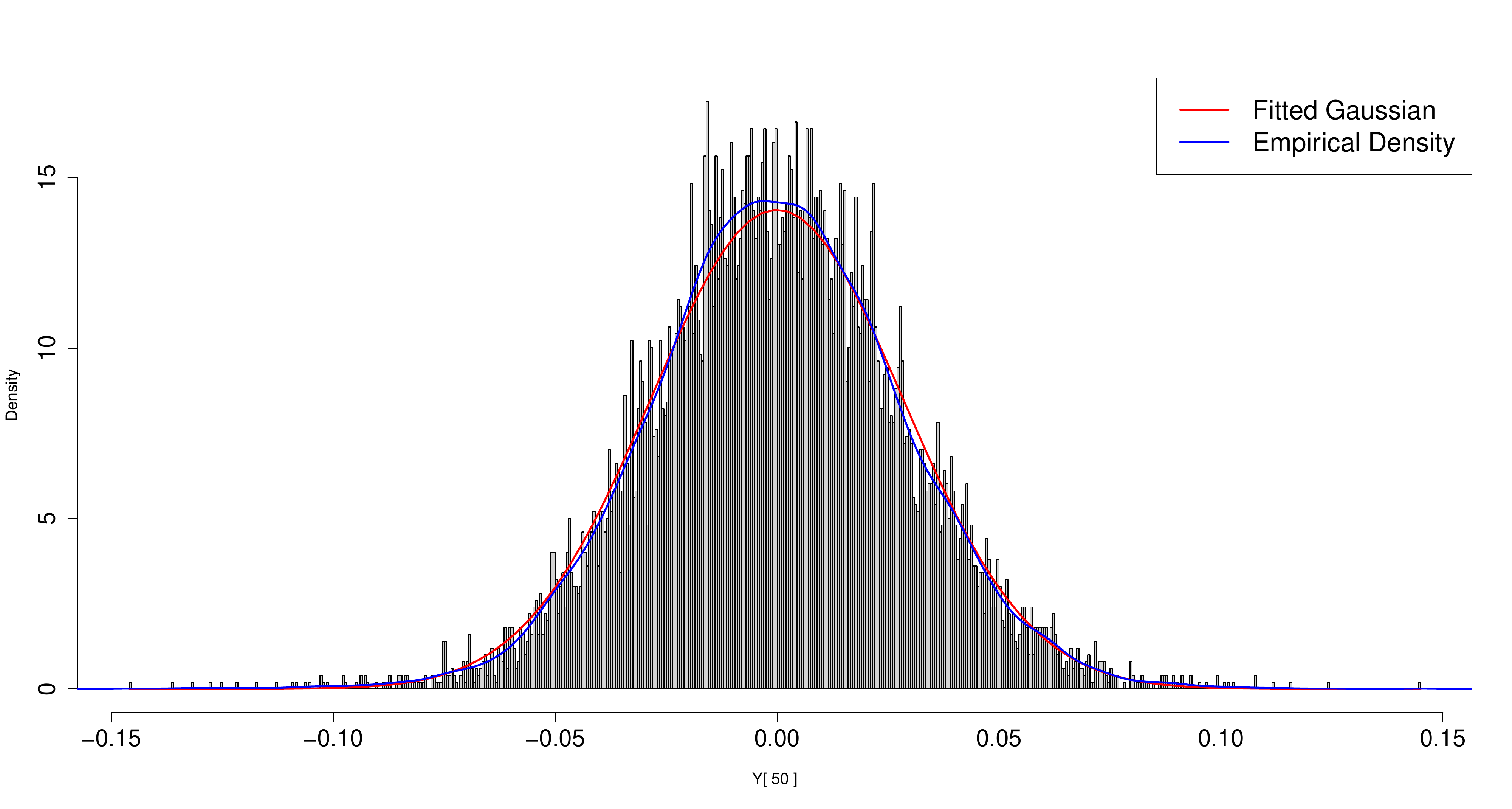}
         \caption{}
         \label{hytn}
     \end{subfigure}
     \caption{histograms of y-positions and empirical density at $t=50$ in the case (a) inviscid, (b) viscous and (c) transport noise, showing different density profiles.}
     \label{hy}
\end{figure}

\begin{figure}[h!]
    \centering
     \begin{subfigure}[H]{0.49\textwidth}
         \centering
         \includegraphics[width=\textwidth]{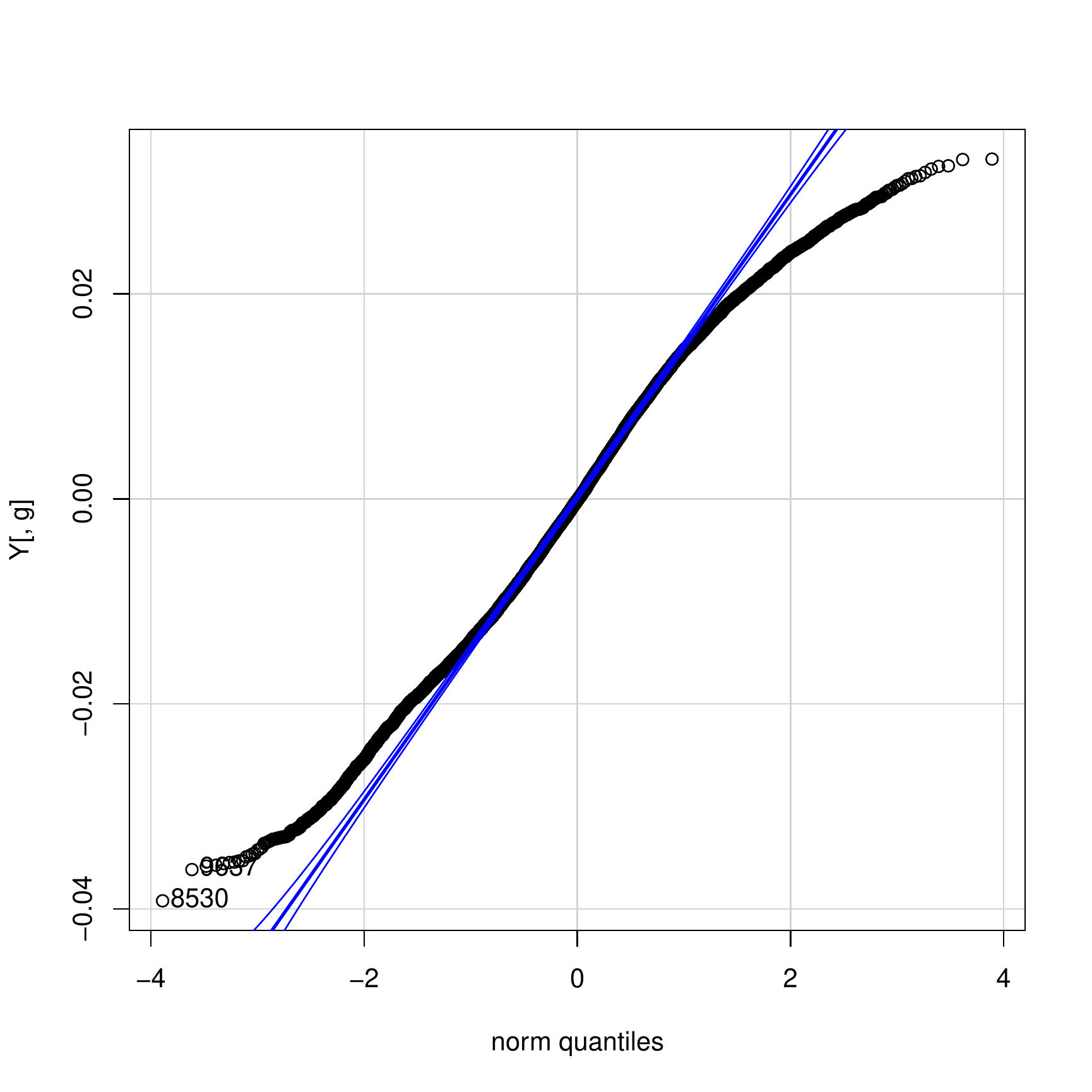}
         \caption{}
         \label{qqd}
     \end{subfigure}
     \begin{subfigure}[H]{0.49\textwidth}
         \centering
         \includegraphics[width=\textwidth]{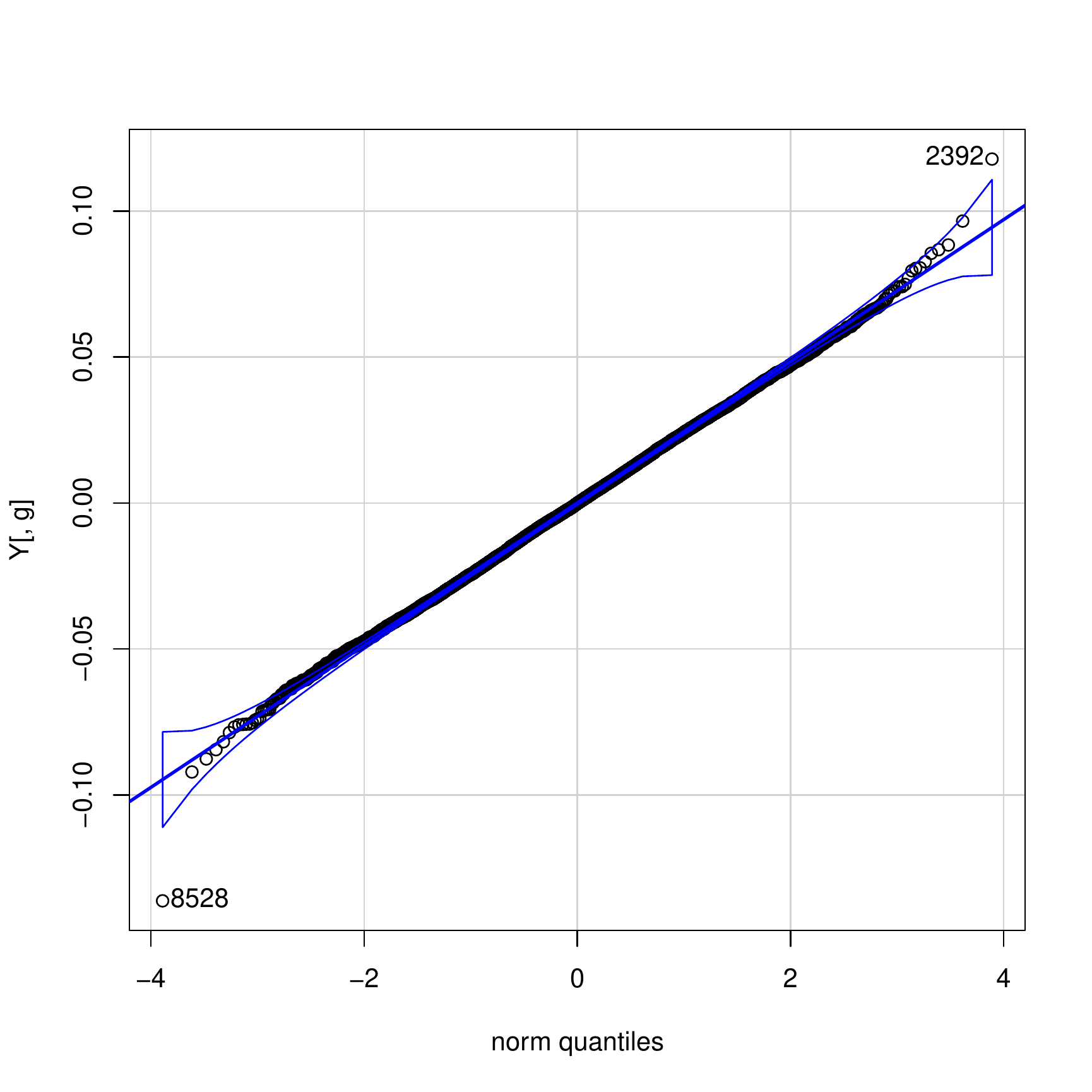}
         \caption{}
         \label{qqn}
     \end{subfigure}
     \begin{subfigure}[H]{0.49\textwidth}
         \centering
         \includegraphics[width=\textwidth]{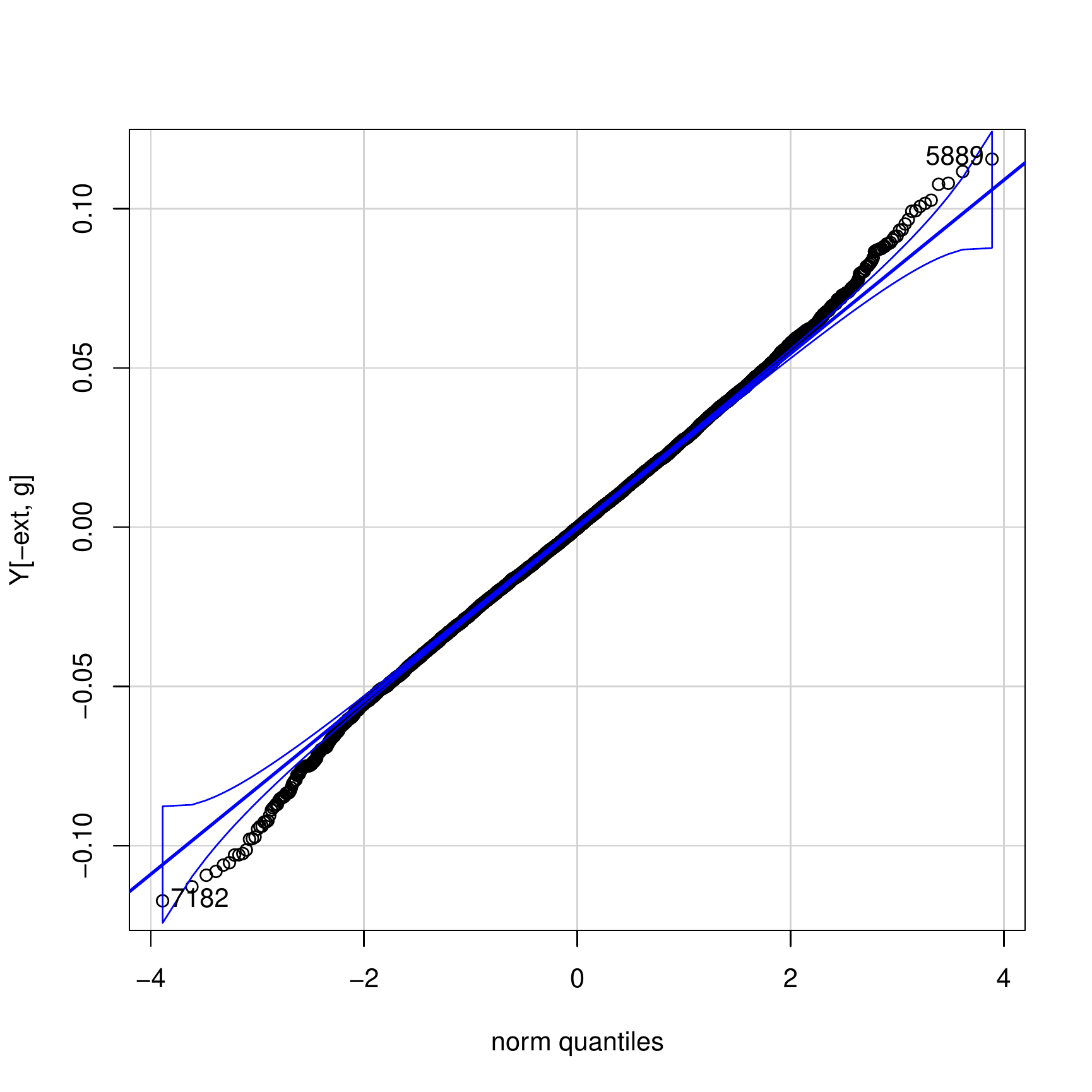}
         \caption{}
         \label{qqtn}
     \end{subfigure}
     \caption{qq-plot of the empirical density in the y-positions, at $t=50$ in the case (a) inviscid, (b) viscous and (c) transport noise.}
     \label{qq}
\end{figure}

Following theoretical results, we know that with our initial condition $\omega_0$, the velocity $u_t$ solves \ref{heq} when $\nu>0$. This is crucial to understanding our system's short-time behaviour and the preservation of the initial configuration. This result states that the empirical density obtained from the y-position, when viscosity is present, maintain a Gaussian profile through time.
In particular, in the case of $\nu=0$, the viscosity follows a classic Euler equation. As such, the y-position profile is far from a Gaussian: it behaves like a multi-modal distribution, concentrated in the proximity of the center of the large structures. This result is supported both by the profile of the particle system and by figures [\ref{hyd},\ref{qqd}]; the qq-plot shows a distinct behaviour for small quantiles. The Kolmogorv-Smirnov test estimates a D-statistic of $0.031$, with a p-value less than $10^{-9}$ confirming the rejection of the Gaussianity hypothesis.

Vice versa, when viscosity is present, i.e. $\nu>0$, as in the case of independent Brownian motions, the noise's diffusive behaviour allows the profile's restoration in the y-direction: the strip configuration is preserved for a longer time.
From the profile of the point vortex system and figures [\ref{hyn},\ref{qqn}], we see that the empirical density approximates well the one of a Gaussian kernel. Moreover, the qq-plot suggests a perfect match with a normal distribution, suggesting the preservation of the strip through time, trading it with more spread on the y-axis. Finally, performing a Kolmogorv-Smirnov test, we see, in fact, a D-statistic of $0.005$ and a p-value greater than $0.9$ suggesting to accept the normality hypothesis. 
This behaviour is preserved throughout the simulation, degrading only at longer times when a few large formations start to rise, as in figure [\ref{fig:three sin x 18n}].

In the transport noise case, we know that we recover the same viscous Euler equation solved in the independent Brownian motion case by applying a particular scaling limit procedure. To this end, we expected that the more stable profile shown in figure [\ref{fig:y equals x17}] presents the same diffusion on the density of the y-position as the case of the independent noise.
This is the case as reported in figures [\ref{hytn},\ref{qqtn}]: the behaviour at a short times, $t=50$, in which the profile still approximates a Gaussian kernel. We perform a qq-plot and a Kolmogorov-Smirnov test on the y-position; we obtain a D-statistic of $0.011$, and a p-value of $0.312$; those results suggest the profile stability and the validity of our hypothesis. 

However, later on ($t=100$), even though the quantity obtained from the KS test is still preserved as in the viscous case, with a D-statistics of $0.008$ and a p-value of $0.48$, the profile degrades as shown in [\ref{fig:three sin x18}]. Those results show that the strip configuration is non-preserved for longer times due to transport noise stretching acting on the point vortices. 
This can be seen in the tail of the distribution in figure [\ref{qqtn}], compared to the viscous case in [\ref{qqn}], which shows at $t=50$ already a different behaviour. This suggests that the environmental noise's effect is not only related to the diffusivity of the strip, but is also responsible for the stretching and formation of different structures. 

\section{Concluding remarks}

In the present work, we have produced numerical simulations of 2D incompressible fluids, also perturbed by transport noise,
using the point vortex method. We focused on the special case of shear flow formation that produced
a Kelvin-Helmholtz instability, in order to test the dissipativity properties of small-space-scale transport noise. We confronted the intrinsic instability generated in the deterministic case with the possible recovery of the stability through
injected noise in the system in the form of transport noise. We showed that, for short times, with a degree less intense than the viscous case,
we can maintain the stability of the strip at the expense of a more small-scale irregularity and diffusion of the profile.

\section{Acknowledgements}
The research of the first author is funded by the European Union (ERC, NoisyFluid, No. 101053472). Views and opinions expressed are however those of the authors only and do not necessarily reflect those of the European Union or the European Research Council. Neither the European Union nor the granting authority can be held responsible for them.


\end{document}